\documentclass[a4paper,11pt]{article}
\pdfoutput=1 

\usepackage{jcappub} 
\usepackage{journal_abbrev}

\usepackage[T1]{fontenc} 
\usepackage{gensymb}
\usepackage[utf8]{inputenc}
\usepackage[T1]{fontenc}
\usepackage{ae,aecompl,graphicx,amsmath,xspace}
\usepackage{esvect}
\usepackage{xcolor,soul,afterpage,subcaption, xparse, xspace, csquotes}
\usepackage[skip=9pt]{caption}
\xspaceaddexceptions{]\}}
\allowdisplaybreaks
\usepackage{tikz}
\usepackage{hyperref}
\usetikzlibrary{shapes}
\usepackage{float}

\DeclareFontFamily{U}{MnSymbolA}{}
\DeclareSymbolFont{MnSyA}{U}{MnSymbolA}{m}{n}
\DeclareFontShape{U}{MnSymbolA}{m}{n}{
    <-6>  MnSymbolA5
   <6-7>  MnSymbolA6
   <7-8>  MnSymbolA7
   <8-9>  MnSymbolA8
   <9-10> MnSymbolA9
  <10-12> MnSymbolA10
  <12->   MnSymbolA12}{}
\DeclareMathSymbol{\rcurvearrowup}{\mathrel}{MnSyA}{185}

\newcommand{\codefont}[1]{{\texttt{#1}}}

\newcommand{\refeq}[1]{Equation~(\ref{eq:#1})}          
\newcommand{\refeqs}[2]{Equations~(\ref{eq:#1})---(\ref{eq:#2})}       
\newcommand{\reffig}[1]{Figure~\ref{fig:#1}}
\newcommand{\refsec}[1]{Section~\ref{sec:#1}}          
\newcommand{\refapp}[1]{Appendix~\ref{app:#1}}
\newcommand{\reftab}[1]{Table~\ref{tab:#1}}

\def\be{\begin{equation}}
\def\ee{\end{equation}}
\def\bea{\begin{eqnarray}}
\def\eea{\end{eqnarray}}

\def\ba#1\ea{\begin{align}#1\end{align}}
\def\bab#1\eab{\begin{equation}\begin{aligned}[b]#1\end{aligned}\end{equation}}

\def\bg#1\eg{\begin{gather}#1\end{gather}}

\newcommand\Mpch{$h^{-1}\;\mathrm{Mpc}$\xspace}

\def\P{\mathcal{P}}

\def\R{\mathcal{R}}

\newcommand{\eps}{\epsilon}

\def\<{\left\langle}
\def\>{\right\rangle}

\newcommand{\vbf}[1]{\mathbf{#1}}

\newcommand{\vx}{\vbf{x}}
\newcommand{\vxhat}{\vbf{\hat{x}}}

\newcommand{\vyhat}{\vbf{\hat{y}}}
\newcommand{\vz}{\vbf{z}}
\newcommand{\vthetahat}{\mathbf{\hat{\theta}}}
\newcommand{\vvarphihat}{\mathbf{\hat{\varphi}}}

\newcommand{\vk}{\vbf{k}}

\newcommand{\vzhat}{\hat{\vz}}

\newcommand{\vnhat}{\vbf{\hat{n}}}

\newcommand{\rhocr}{\rho_{\mathrm{cr}}}

\newcommand{\Om}{\Omega_\mathrm{m}}

\def\rnd{\mathrm{rnd}}
\def\obs{\mathrm{obs}}

\def\emph#1{\textit{#1}}

\definecolor{nicegreen}{HTML}{2CA02C}
\definecolor{darkgreen}{rgb}{0.0,0.5,0.0}

\usepackage{todonotes}
\newcounter{todocounter}

\title{\boldmath Field-level inference of galaxy intrinsic alignment from the SDSS-III BOSS survey}

\author[a]{Eleni Tsaprazi,}
\author[b]{Nhat-Minh Nguyen,}
\author[a,c]{Jens Jasche,}
\author[b]{Fabian~Schmidt}
\author[c]{and Guilhem Lavaux}

\affiliation[a]{The Oskar Klein Centre, Department of Physics, Stockholm University, Albanova University Center, SE 106 91 Stockholm, Sweden}
\affiliation[b]{Max–Planck–Institut f\"ur Astrophysik, Karl–Schwarzschild–Straße 1, D–85748 Garching, Germany}
\affiliation[c]{Sorbonne Université, CNRS, UMR 7095, Institut d'Astrophysique de Paris, 98 bis bd Arago, 75014 Paris, France}

\emailAdd{eleni.tsaprazi@fysik.su.se}
\emailAdd{minh@mpa-garching.mpg.de}
\emailAdd{jens.jasche@fysik.su.se}
\emailAdd{fabians@mpa-garching.mpg.de}
\emailAdd{guilhem.lavaux@iap.fr}

\abstract{As a large-scale overdensity collapses, it affects the orientation and shape of galaxies that form, by exerting tidal shear along their axes. Therefore, the shapes of elliptical galaxies align with the tidal field of cosmic structures. This intrinsic alignment provides insights into galaxy formation and the primordial universe, complements late-time cosmological probes and constitutes a significant systematic effect for weak gravitational lensing observations. In the present study, we provide constraints on the linear alignment model using a fully Bayesian field-level approach, using galaxy shape measurements from the SDSS-III BOSS LOWZ sample and three-dimensional tidal fields constrained with the LOWZ and CMASS galaxy samples of the SDSS-III BOSS survey. We find 4$\sigma$ evidence of intrinsic alignment, with an amplitude of $A_I=2.9 \pm 0.7$ at 20\Mpch.
}

\begin{document}
\label{firstpage}
\maketitle

\section{Introduction}
\label{sec:intro}
Galaxy formation typically follows the gravitational collapse of large-scale overdensities. During the collapse, gravitational tidal forces lead to anisotropies in the shape and orientation of galaxies. Therefore, the shape and orientation of galaxies carries information on the tidal field at their location and over the course of their formation history. Growing evidence over the last two decades suggests that the intrinsic orientation of elliptical galaxies is indeed subject to a large-scale coherent alignment, often referred to as \emph{intrinsic alignment} \citep[e.g.][]{2002MNRAS.333..501B, Mandelbaum:2006, Hirata:2007, Joachimi:2011, Blazek:2011, Singh:2015, 2022MNRAS.509.3868H}. This large-scale correlation is attributed to the alignment of galaxy shapes for pressure-supported elliptical galaxies (tidal stretching) and that of angular momenta in rotation-supported, spiral galaxies (tidal torquing) with the large-scale tidal shear \citep[e.g.][]{2001MNRAS.320L...7C,Hirata:2007,2019PhRvD.100j3506B}. As weak lensing analyses infer the total projected mass from projected galaxy shapes, intrinsic shape alignments can interfere with weak lensing measurements, which are based on the assumption that intrinsic galaxy shapes are uncorrelated. It is now widely recognized that intrinsic alignment is a non-negligible source of contamination for weak lensing surveys. Intrinsic alignment contamination can strongly bias cosmological conclusions drawn from weak lensing observations \citep[e.g.][]{2015PhR...558....1T}. Therefore, mitigating it is necessary. This statement applies to existing as well as future surveys, like the Kilo-Degree Survey, Dark Energy Survey, Hyper-Suprime Cam, Euclid, the Legacy Survey of Space and Time and the Square Kilometer Array \citep{Euclid:2013, DES:2016, HSC:2018, LSST:2019, SKA:2020,KiDS:2020}. Further, intrinsic alignment is a probe of specific types of primordial non-Gaussianity \citep{2015PhR...558....1T,Schmidt:2015,chisari/etal:2016,kogai/etal}, redshift-space distortions, Baryon Acoustic Oscillations \citep{Chisari:2013} and the stochastic gravitational wave background \citep{2012PhRvD..86h3513S,chisari/etal:2014}.

The evolution of intrinsic alignment in time can constrain the response of galaxy shapes to external tidal fields. While galaxy shapes are nonlinear functions of their environment and tidal shear
\citep[e.g.][]{Larsen:2016,2009IJMPD..18..173S,1955MNRAS.115....3S,1969ApJ...155..393P,1970Ap......6..320D,1984ApJ...286...38W},
a first-order expansion leads to the linear alignment model which is expected to be valid on large scales \citep{Catelan:2001,2001MNRAS.320L...7C,Hirata:2004}; specifically, scales larger than the Lagrangian radius of galaxy host halos and the scale where large-scale structure becomes nonlinear, i.e. $R\gtrsim 20-30$\Mpch{} \citep{vlah/etal, 2015JCAP...08..015B}.

The linear alignment model has been extensively applied to observational analyses. Past works have found significant correlations of intrinsic galaxy shapes on scales up to 60\Mpch{} \citep{Mandelbaum:2006,Hirata:2007,Blazek:2011,Joachimi:2011,Singh:2015}.
Potential luminosity evolution of linear alignment is a topic of active research \citep[e.g.][]{Singh:2015,2016MNRAS.461.2702C}, which suggests a trend of stronger alignments for brighter galaxies. Linear alignment is also found to be stronger for redder galaxies than bluer ones, but relatively constant within red samples \citep[e.g.][]{Joachimi:2011,2013MNRAS.436..819J,2013MNRAS.432.2433H,Singh:2015,2019A&A...624A..30J,2021MNRAS.508..637S}. This trend has also been observed in a range of intrinsic alignment models in recent analyses \citep[e.g.][]{2019MNRAS.489.5453S}.

Previous studies constrained intrinsic alignment models by compressing galaxy shape catalogs to shape-shape and shape-position correlations \citep[e.g.][]{Joachimi:2011, Blazek:2011, Singh:2015}. In this paper, we instead propose a field-level approach. We infer the linear alignment amplitude directly from the cross-correlation of luminous red galaxy shapes in the LOWZ sample of the SDSS-III BOSS DR11 survey with the local, large-scale tidal shear at the location of each galaxy.

First, we obtain our tidal shear estimates from an ensemble of dark matter density fields, constrained with SDSS-III BOSS galaxy data using the \codefont{BORG} algorithm \citep{2010MNRAS.406...60J,Jasche:2013,2015MNRAS.447.1204J,Lavaux:2019,Jasche:2019}. Then, we cross-correlate these estimates with the observed galaxy shapes and constrain the linear alignment model on scales of $\geq15.6$\Mpch{}, within the redshift range of $0.21<z<0.36$.
The lower limit is the resolution of the density fields in our study. By construction, the \codefont{BORG} density resimulations incorporate the full statistics of the large-scale structure, while accounting for uncertainties in the galaxy data. 
In the present study, we use the LOWZ galaxy ellipticity catalog of the SDSS-III BOSS DR11 survey  \citep{Reyes:2012,2013AJ....146...32S,2015ApJS..219...12A,2006AJ....131.2332G,2013AJ....145...10D,2011AJ....142...72E,Singh:2015}. As our analysis involves the tidal field at mildly nonlinear scales and the latter was evolved using 2LPT, the expression ``linear alignment'' that is used here, typically corresponds in the literature to the ``nonlinear alignment (NLA) model''.

Throughout our analysis, we assume $H_0 = 67.74\;\mathrm{km\,s^{-1}\,Mpc^{-1}}$, $\Omega_{m} = 0.2889$, $\Omega_b = 0.048597$, $\Omega_r = 0$, $\Omega_k = 0$, $\Omega_\Lambda = 0.7111$, $w = -1$, $n_S = 0.9667$, $\sigma_8 = 0.8159$, consistent with the parameters assumed in the density field inference \citep{Lavaux:2019}. The paper is structured as follows. In \refsec{LA_model}, we review the linear alignment model and describe how we relate our constrained tidal field to the linear alignment prediction. In \refsec{data_model}, we present our data model for galaxy shapes. In \refsec{tidal_data}, we describe the large-scale dark matter density, tidal field and shape data. In \refsec{posterior}, we describe our Bayesian inference framework and our method to sample from the posterior of the alignment amplitude. In \refsec{results} and \refsec{discussion}, we report and discuss our results, respectively.

\section{The linear intrinsic alignment model}\label{sec:LA_model}

Elliptical galaxies can be stretched or squeezed under the influence of external tidal fields \citep[e.g.][]{Catelan:2001}. According to the linear alignment model, this contribution to galaxy shapes is linearly proportional to the tidal field at the location of galaxies. Here, we determine the linear alignment amplitude by correlating measured components of galaxy shapes (ellipticities), $e^I$, and the local gravitational tidal shear tensor
\be
    T_{ij}\equiv\frac{\partial^2\Phi}{\partial x_i \partial x_j},
\ee
where $\{ x_i : i \in \{1,2,3\}\}$ are comoving Cartesian coordinates and $\Phi$ the gravitational potential. Galaxy ellipticities are two-dimensional, while the tidal tensor is three-dimensional. In what follows, three-dimensional spatial vectors are denoted by $\vx$. We label galaxies and tidal tensor samples using the subscripts $g$ and $s$, respectively. $\<X\>$ denotes an ensemble average of $X$ over tidal tensor samples, while $\bar{X}$ denotes a spatial average. Curly brackets indicate ensemble over the bracketed quantity's index. We therefore project the latter, at the location of each galaxy, onto a two-dimensional surface perpendicular to the line of sight, $\vnhat$. We select an orthonormal basis $\left(\vnhat,\mathbf{\hat{\theta}},\mathbf{\hat{\varphi}}\right)$ such that
\bea
\vthetahat &=& \cos{\theta}\cos{\varphi}\,\vxhat+ \cos{\theta}\sin{\varphi}\,\vyhat - \sin{\theta}\,\vzhat,\\
\vvarphihat &=& -\sin{\varphi}\,\vxhat+\cos{\varphi}\,\vyhat,
\label{eq:thetahat_varphihat}
\eea
with $\left(\vxhat,\,\vyhat,\,\vzhat\right)$ being the orthonormal basis of the Cartesian coordinates, $\theta$ the declination and $\varphi$ the right ascension of the galaxy. We then decompose the local tidal tensor into \citep{Schmidt:2012, Schmidt:2015}
\be
T_\pm = \sum_{i=1}^3\sum_{j=1}^3 {m^i}_{\mp} {m^j}_{\mp} T_{ij},
\label{eq:decomposition}
\ee
where $T_\pm\equiv T_1\pm \iota T_2$ are complex spin-$\pm$2 fields. Under a rotation by $\psi$ around the line of sight $\vnhat$, the two fields  $T_{\pm}$ transform as $T'_\pm(\vnhat) \to T'_\pm(\vnhat) = e^{\iota2\psi}m_\pm(\vnhat)$ and $m_\pm^i \equiv \vvarphihat^i\mp \iota\vthetahat^i/\sqrt{2}$, $\iota$ being the imaginary unit $\sqrt{-1} \equiv \iota$. The two fields $m_{\pm}$ transform as spin-$\pm$1 fields, i.e. $m'_\pm(\vnhat) \to m'_\pm(\vnhat) = e^{\iota\psi}m_\pm(\vnhat)$.
The resulting quantities $T_\pm$ are the components of the trace-free part of $T_{ij}$ projected on the plane that is perpendicular to the line of sight to each galaxy. 
In the ellipticity catalog that we use, $e^{\obs}_1$ and $e^{\obs}_2$ are defined as ellipticity components along the right ascension $\vvarphihat$ and declination $\vthetahat$ respectively, such that positive $e^{\obs}_1$ correspond to east - west elongation and positive $e^{\obs}_2$ to northeast - southwest elongation \citep{Reyes:2012}. We adopt the same convention. Upon the decomposition, we end up with projected tidal fields at the locations of galaxies, which are different for every tidal field realization.

Due to the correlation length of the tidal field, the linear alignment signal can persist up to scales much larger than those relevant to galaxy formation.
In order to probe linear alignment on a range of scales, we estimate the smoothed tidal tensor from the dark matter density contrast, $\delta$, in Fourier representation for different smoothing scales, $R$, through the Poisson equation 
\be
T_{ij}(\vk,R,z) = \frac{4\pi G\rho_\mathrm{cr}\Omega_{m,0}}{a(z)}\,\left[\frac{k_i k_j}{|\vk|^2}\,W_\mathrm{SK}(\mathbf{k}, R)\,\delta(\vk,z)\right],
\label{eq:tidal_tensor}
\ee
where $\vk$ is the wavevector, $\rho_\mathrm{cr}$ the critical density of the Universe, $\Omega_{m,0}$ the matter density parameter today, $a=1$ the cosmic scale factor and $z$ the redshift of each galaxy. $W_\mathrm{SK}$ represents the tophat filter with which we choose to smooth the density field and is given by
\be
W_\mathrm{SK}(\mathbf{k},R) = \begin{cases} 1 & \text{if } |\mathbf{k}|\leqslant1/R \\
0 & \text{otherwise.}      
\end{cases}
\ee
We will refer to this filter as `sharp-k', derived from a sharp cut of Fourier modes whose magnitudes are above the cut-off $1/R$. We use it to prevent the contamination of a given scale with information from smaller scales, where unmodeled nonlinear processes could bias our signal. By additionally filtering the tidal field with the sharp-k filter, we obtain estimates of the tidal field that retain information above the cut-off scale $R$.

Then, the ellipticity of a galaxy $e^\mathrm{I}$ at position $\mathbf{x}$ can be written as a function of the tidal field
\be 
e^\mathrm{I}_1 \pm \iota\, e^\mathrm{I}_2 = -\frac{C}{4\pi G}\,T_\pm,
\label{eq:LA_model}
\ee
where $e^\mathrm{I}_1$, $e^\mathrm{I}_2$ denote ellipticities measured along the galaxies' right ascension and declination. $G$ is the gravitational constant and $C$ the linear alignment amplitude. We note that in our normalization convention the intrinsic alignment amplitude is taken to be at the time of observation. In the flat-sky limit and fixing the line of sight to the $z$ axis,
\refeq{LA_model} reduces to the relation
\be
e^\mathrm{I}\equiv\left[e^\mathrm{I}_1,\,e^\mathrm{I}_2\right] = -\frac{C}{4\pi G}\left[\partial_x^2-\partial_y^2,\,2\partial_x\partial_y\right]\,\Phi(\mathbf{x}).
\label{eq:LA_model_Catelan} 
\ee

\section{The data model}\label{sec:data_model}

The observed galaxy ellipticities, $\mathbf{e}_\alpha^{\obs}$, receive contributions from processes other than linear alignment, which at lowest order are additive: 
\be
\mathbf{e}_\alpha^{\obs} = \mathbf{e}_\alpha^\mathrm{I} + \mathbf{\epsilon}_\alpha^\mathrm{msm} + \mathbf{\epsilon}_\alpha^\mathrm{WL} + \mathbf{\epsilon}_\alpha^{\rnd},
\label{eq:observed_e}
\ee
where $\alpha \in \{1,2\}$ is the index for the two ellipticity components, $\epsilon^{\mathrm{msm}}$ is the shape measurement error and $\mathbf{\epsilon}^\mathrm{WL}$ is due to weak lensing from foreground matter. The median redshift of our sample is $\sim 0.3$. The expected contribution from weak lensing to the observed ellipticity is $\langle {(\mathbf{\epsilon}^\mathrm{WL})}^2 \rangle \lesssim 10^{-4}$ \citep[][Eq. 42]{2002MNRAS.335..909B}.
Therefore, the typical lensing contribution to the ellipticity components is less than $\sim 0.01$, whereas the mean standard deviation of the intrinsic alignment contribution to galaxy shapes in our inference is $\sim 0.07$. 
We therefore choose to ignore the weak lensing contribution. When applying this method to higher-redshift sources, lensing could be accounted for by integrating the inferred matter distribution in the inference along the line of sight to each galaxy. $\epsilon^{\rnd}$ is a stochastic contribution to galaxy shapes, which captures small-scale physical alignment processes that are not considered in the linear alignment model.

Estimating the random noise as the standard deviation of $e^{\obs}$, we find that $\epsilon^{\rnd}$ is at least one order of magnitude larger than the shape measurement uncertainties.
We therefore assume that the total ellipticity uncertainty is dominated by random noise, leading to a variance that is universal across our galaxy sample. Our single-galaxy data model then reads
\be
e_\alpha^{\obs} = -\frac{C(R)}{4\pi G} T_\pm(R) + \epsilon_\alpha^\mathrm{rnd}(R),
\label{eq:datamodel}
\ee
where $R$ is the scale of the smoothed tidal field and $\epsilon_\alpha^\mathrm{rnd}$ is drawn from a Gaussian distribution with a variance that only depends on $R$. We assume that the random component is uncorrelated between different galaxies. To simplify notations, we will refer to $C$ as the amplitude at a given smoothing scale. Our smoothing of the tidal field $T_{\pm}$ implies that the random noise component $\epsilon^\rnd$ becomes scale-dependent. We also expect this scale dependence to arise from small-scale physics, which would bias our inference non-trivially on scales smaller than the original resolution of the tidal fields.

While galaxy shapes are measured at the scale of a single galaxy, we smooth the tidal fields derived from the inferences at a larger scale. This introduces an unphysical scale dependence to the linear alignment amplitude when the smoothing scale $R$ is not much larger than the inference resolution.
In order to compare our results for the amplitude of the linear alignment  to previous literature results, we correct $C(R)$ for the smoothing kernel which is implicitly introduced in the original inference. We describe our corrections in Appendix~\ref{app:smoothing_correction}. 

\section{Data}\label{sec:tidal_data} 

In this section, we describe the datasets that enter the right- and left-hand side of our data model in \refeq{datamodel}. Those include, respectively the tidal field components $T_\pm(R)$ (\refsec{SDSS-III_recon}) for the right-hand side and the observed galaxy shapes $e^{\obs}_\alpha$ (\refsec{image_data}) for the left-hand side.

\subsection{The large-scale tidal shear}\label{sec:SDSS-III_recon}

In order to derive the $T_\pm(R)$ components of the tidal field at the location of each galaxy, as seen from \refeqs{decomposition}{tidal_tensor}, we need the dark matter density field, $\delta(k)$. In this work, the latter is supplied by the density resimulations of the SDSS-III BOSS volume with a gravity-only model. The resimulation is based on an inference of initial conditions constrained with the SDSS-III LOWZ and CMASS galaxy data \citep{2013AJ....145...10D,Lavaux:2019} by the \codefont{BORG} algorithm \citep{2010MNRAS.406...60J,Jasche:2013,2015MNRAS.447.1204J,Lavaux:2019,Jasche:2019}. We choose to include neither light-cone effects nor redshift-space distortions, as these effects would significantly complicate the conversion from density to rest-frame tidal field. Given the narrow redshift range of the volume used here, light-cone effects are expected to be negligible. Neglecting the displacement into redshift space will have an impact on comoving scales comparable to the typical displacement, which is less than $20$\Mpch, while we will focus on scales $R \gtrsim 20$\Mpch below.

\codefont{BORG} performs a Bayesian inference of the initial conditions within a given region of the observed universe, constrained by galaxy observations and survey characteristics. To achieve this task, it assumes a model of structure formation and galaxy bias. The former is used to evolve initial conditions in time. The latter allows \codefont{BORG} to compare the present-day density fields with the galaxy observations. The density field posterior is approximated by Markov Chain Monte Carlo (MCMC) samples, which incorporate all observational uncertainties. This process results in statistically plausible samples of the present-day, three-dimensional dark matter density field, as traced by the input galaxy survey. The MCMC samples provide a full statistical description of the complex dark matter distribution. 

Provided with the inferred initial conditions of the SDSS-III BOSS volume \citep{Lavaux:2019}, we resimulate the present-day matter density field in comoving rest-frame using second-order Lagrangian Perturbation Theory. The resolution of the density and tidal fields is $\sim 15.6$\Mpch{} ($\sim 1\degree$ at the median redshift of the ellipticity sample). The inference domain is a box of side length $L = 4000$\Mpch{} and grid resolution $N=256$. In this study, we estimate the tidal fields from 837 MCMC \codefont{BORG} density resimulations, separated by 10 MCMC steps, extracted from the original chain \citep{Lavaux:2019}. Our results are robust to the number of density field realizations used.

We use the tidal fields inferred at redshift $z=0$. We multiply the projected ellipticity of each galaxy, $T_\pm$, by $B(z) = (1+z)D(z)$, $D(z)$ being the linear growth factor, according to \refeq{tidal_tensor}, in order to account for the redshift dependence of the intrinsic alignment amplitude. We adopt the instantaneous alignment scenario \citep[e.g.][]{2015JCAP...08..015B,2016MNRAS.456..207K} to report our results and the primordial alignment scenario to compare them to the findings by \citep{Singh:2015}. We verify the validity of this step as follows. With the above procedure, we obtain, for a single tidal field realization, an intrinsic alignment amplitude of $2.3\pm0.6$. We then resimulate the tidal field for this realization using 2LPT to the median redshift of the sample, $z=0.26$. We then derive the intrinsic alignment amplitude at $z=0.26$ and find $2.3\pm0.6$. We thus confirm that $B(z)$ captures the redshift-dependence of the intrinsic alignment signal. This is expected, since on the scales probed, linear growth is a good approximation.

The dark matter density field, both in the \codefont{BORG} inference \citep{Lavaux:2019} and our resimulation, is convolved with a cloud-in-cell (CIC) kernel by construction, resulting in a resolution of $R_0 = 15.6$\Mpch{}. In order to probe the linear alignment amplitude on various scales, we smooth all tidal fields by convolving them with a sharp-k filter (cf. \refeq{tidal_tensor}).
The smoothing scales that we consider are $R\leq120$\Mpch{} ($\leq 8\degree$ at the median redshift of the ellipticity sample). The resulting tidal field posterior consists of 837 tidal field realizations per smoothing scale. We then derive the projected components of the tidal shear at the location of a given galaxy from the smoothed tidal fields using the Nearest Grid Point (NGP) assignment scheme. We first transform each galaxy's equatorial coordinates and redshift to comoving Cartesian coordinates, $x_i$. We then convert the Cartesian coordinates to grid indices, $l_i$, ($i \in \{1,2,3\}$), as follows
\be
 l_i = \Big\lfloor{\frac{x_i}{R_0} + 0.5\Big\rfloor},
\ee
where $\lfloor{\phantom{a}\rfloor}$ represents the floor function. 

\subsection{Galaxy shapes}\label{sec:image_data}

The galaxy shape data that we use is part of the original catalog constructed using the LOWZ sample of the SDSS-III BOSS survey \citep{Reyes:2012,Singh:2015}. The sample consists of luminous red galaxies and covers the redshift range $0.16 < z < 0.36$. We apply a lower redshift cutoff at $z=0.21$, corresponding to the lower redshift limit of the inference volume \citep{Lavaux:2019}. We remove galaxies outside the SDSS-III BOSS survey mask. We additionally remove galaxies outside of the LOWZ and CMASS footprints assumed by the \codefont{BORG} inference. Since the latter covers more sky area, some galaxies outside of the LOWZ footprint remain after the cut. Our resulting sample consists of 142~514 galaxies, with ellipticity components in the range $[-2,2]$. 

\section{The intrinsic alignment posterior}\label{sec:posterior}

In this section, we present our statistical framework for the linear alignment amplitude inference. Unless specified otherwise, we write $e^{\obs}$ as $e$ for readability. The subscripts $g$ and $s$ denote the galaxy and tidal field sample indices, respectively. We omit these indices wherever we refer to the ensemble mean over the respective quantity.
Recall that the reason for considering the tidal field sample indices is that the \codefont{BORG} algorithm provides an ensemble of plausible realizations of the large-scale structure posterior.
Our inference framework considers both ellipticity components jointly. We assume that all galaxy ellipticity measurements have the same variance, $\sigma^2$, dominated by random processes (see \refsec{data_model}). Note that the formulation below does not consider the smoothing correction, described in Appendix \ref{app:smoothing_correction}. The correction is applied later to the final linear alignment amplitude and variance posteriors.

We then write the likelihood for the observed galaxy shapes given a single realization of the tidal field, assuming that the random noise on the right-hand side of \refeq{datamodel} is Gaussian-distributed, as
\be
   \P\left(e \Big|C, T_{s}, \sigma^2 \right) = \prod_{g=1}^{2N_g}
   \frac{1}{\sqrt{2 \pi \sigma^2}} \exp\left[-\frac{\left (e_{g} - C T_{g,s} \right)^2}{2\sigma^2}\right].
   \label{eq:likelihood_singlesample}
\ee
Note that $e_g$ consists of both ellipticity components, which we assume to be affected by the same variance. Therefore, we treat $e_1$ and $e_2$ as independent measurements, and the product in \refeq{likelihood_singlesample} runs to $2N_g$, as every galaxy is associated with two ellipticity measurements. Further, notice the dependence of the intrinsic alignment amplitude on the tidal field realizations; this derives from the fact that every realization is a plausible tidal field sample given the SDSS-III BOSS survey.

The choice of a Gaussian likelihood for galaxy shapes is a strongly simplifying assumption. It is only accurate if galaxy intrinsic shapes are uncorrelated (apart from the intrinsic alignment contribution which is modeled explicitly) and if the effective number of galaxies entering the tidal alignment estimate is sufficiently large, i.e. if each $e_g$ here was actually estimated from a large sample of independent galaxy shapes. The first assumption is likely to be fairly accurate on the scales considered here. In order to estimate the effect of the latter assumption, consider the case where $T_{g,s}$ is approximately the same for all galaxies $g$. Then, we attempt to infer the mean shape $C T_{g,s}$ from a large set of noisy individual shapes. \refeq{likelihood_singlesample} shows that the maximum-likelihood estimate of $C$ is essentially the sample mean of $e_{g}$, divided by the constant $T_{g,s}$. This is unbiased in the large-$N_g$ limit as long as the $e_{g}$ are independent and identically distributed, even for a non-Gaussian distribution of $e_{g}$. In practice, $T_{g,s}$ is not constant, but varies from galaxy to galaxy (albeit slowly, since we filter $T_{g,s}$ on fairly large scales). However, this provides a qualitative justification for why we expect the galaxy shape likelihood not to bias our estimator significantly. 

Our posterior predictive check (\refapp{ppc}) indicates that this choice is likely suboptimal, due to the presence of heavy tails in the observed shape distribution. A naive application of the Gaussian likelihood, with a fixed variance $\sigma^2$ obtained from the standard deviation of the data distribution, is therefore sensitive to outliers (see also the discussion in Section 4.3.2 of \citep{Reyes:2012}).
Here we mitigate this sensitivity by jointly inferring the variance (alongside with the linear alignment amplitude) -- essentially allowing for a larger variance than that of the data distribution. \reffig{ppc} shows that, indeed, the variance of our posterior is larger than that of the data distribution.

Assuming a uniform prior on the linear alignment amplitude, we derive the linear alignment amplitude posterior marginalized over the tidal field realizations (see \refapp{posterior_formulation}, analogous to the derivation in \citep{Nguyen:2020a})
\be
\P\left(C \Big|e, \sigma^2\right) \propto   \sum_{s=1}^{N_s}  \frac{\lambda_{s}}{\sqrt{2\pi\,\nu_{s}^2 }}\exp\left[-\frac{\left(C-\mu_{s}\right)^2}{2\nu_{s}^2}\right],
\label{eq:amplitude_posterior_2}
\ee
which is a Gaussian mixture distribution. Each Gaussian component is associated with a realization of the tidal field. The mixture weights, $\lambda_{s}$, are given by (\refapp{posterior_formulation}):
\be
\lambda_{s} = \frac{\exp\left[\omega_{s} \, + \frac{1}{2}\mathrm{ln}\left(2\pi\,\nu_{s}^2\right)\right]}{ \sum_{s=1}^{N_s} \exp\left[\omega_{s} \, + \frac{1}{2}\mathrm{ln}\left(2\pi\,\nu_{s}^2\right)\right]},
\label{eq:lambda_s}
\ee
with
\be
\nu_{s}^{-2} = \sum_{g=1}^{2N_g} \frac{T_{g,s}^2}{\sigma_g^2}
\label{eq:nu2_s}
\quad\text{,}\quad 
\omega_{s} = \frac{\mu_{s}^2}{2\nu^2_{s}} 
\quad\text{,}\quad 
\mu_{s} = \frac{\sum_{g=1}^{2N_g} e_{g} T_{g,s} / \sigma_g^2}{\sum_{g=1}^{2N_g} T_{g,s}^2/\sigma_g^2}
\ee
\refeq{amplitude_posterior_2} is a Blackwell-Rao estimator \cite{Blackwell:1947, Rao:1992}. The mixture weights, $\lambda_{s}$, assign preference to realizations of the tidal field that are more compatible with the observed galaxy shapes.

Assuming that the random ellipticity noise is Gaussian-distributed and universal across the galaxy sample, we perform a joint inference of the linear alignment amplitude and the ellipticity variance. We assume an inverse-gamma (IG) conjugate prior on $\sigma^2$ of the form $P(\sigma^2|C, T_{g,s})\sim\mathrm{IG}(\theta_1,\theta_2)$, where $\theta_1,\theta_2$ are the shape and scale parameters of the distribution respectively. We choose this prior as it is close to a Jeffrey's prior and yields a bonafide distribution, namely a distribution that integrates to unity. Then, we write the conditional posterior for the ellipticity variance as
\be
P(\sigma^2|C, e) \propto \sum_{s=1}^{N_s} \mathrm{IG}\left(\sigma^2\Bigg|\theta_1+\frac{N_g}{2};\; \theta_2+\frac{1}{2}\sum_{g=1}^{2N_g}(e_{g} - C{T_{g,s}})^2\right),
\label{eq:variance_posterior_2}
\ee
where
\be
\mathrm{IG}(\sigma^2|\theta_1,\theta_2)=\frac{\theta_2^{\theta_1}}{\Gamma(\theta_1)}x^{-\theta_1-1}\exp{\left(-\frac{\theta_2}{\sigma^2}\right)},
\ee
$\Gamma(\sigma^2)$ being the gamma function.

In order to sample from the posterior distributions of \refeq{amplitude_posterior_2} and \refeq{variance_posterior_2}, we adopt a Gibbs sampling approach \citep{4767596, 10.2307/2289776}. As a first step, we sample from the conditional variance posterior, namely the random noise posterior. In the second step, we sample from the linear alignment amplitude posterior, conditional on the previous variance sample. We then iterate this process for $N=10,000$ sampling steps. We illustrate our block-sampling approach, for a single tidal field realization as
\begin{eqnarray}
  \sigma^2 &\curvearrowleft&  \nonumber  P\left(\sigma^2|C, e, T_{s}\right) \nonumber\\ 
  &\phantom{\sigma^2 \curvearrowleft}& \nonumber\\ 
  C &\curvearrowleft& \P\left(C\Big| \sigma^2, e, T_{s}\right). 
  \label{eq:Gibbs}
\end{eqnarray}
The curved arrows, $\curvearrowleft$, indicate random draws. \refeq{Gibbs} illustrates the iterative process in which we first draw a variance sample and then use it to draw an intrinsic alignment amplitude sample. In order to be maximally agnostic we set the variance prior to $\mathrm{IG}(10^{-3},10^{-3})$. While the Jeffrey's prior in our case is $\mathrm{IG}(0,0)$, we choose small nonzero values of the IG parameters for numerical reasons. This is a valid approximation as long as the variance is much greater than $10^{-3}$ and we show that this is the case in \refsec{results}. The dependence of our results on the IG prior is negligible. We further test the convergence behavior of our approach using a Gelman - Rubin test, mock data tests and a comparison to the analytical expectation of the posterior means to the limit of a fixed uncertainty. All tests and the sensitivity of our inference to the choice of prior are discussed in Appendix \ref{app:validation}. In principle, one could directly sample from the mixture distribution in \refeq{amplitude_posterior_2}, considering the individual weights, $\lambda_s$. In order to account for all realizations, we choose to rather sample from the Gaussian distributions corresponding to every tidal field realization separately and combine the posteriors (\refapp{posterior_formulation}) for computational convenience.

\section{Results}\label{sec:results}

We now present our results of the joint linear alignment amplitude and ellipticity variance inference. The posterior moments presented here are the result of running $N_s=837$ independent sampling chains, each consisting of 10~000 MCMC steps. Both ellipticity components are sampled in each chain, in order to obtain combined constraints.

\begin{figure}
    \centering
    \includegraphics[width=\textwidth]{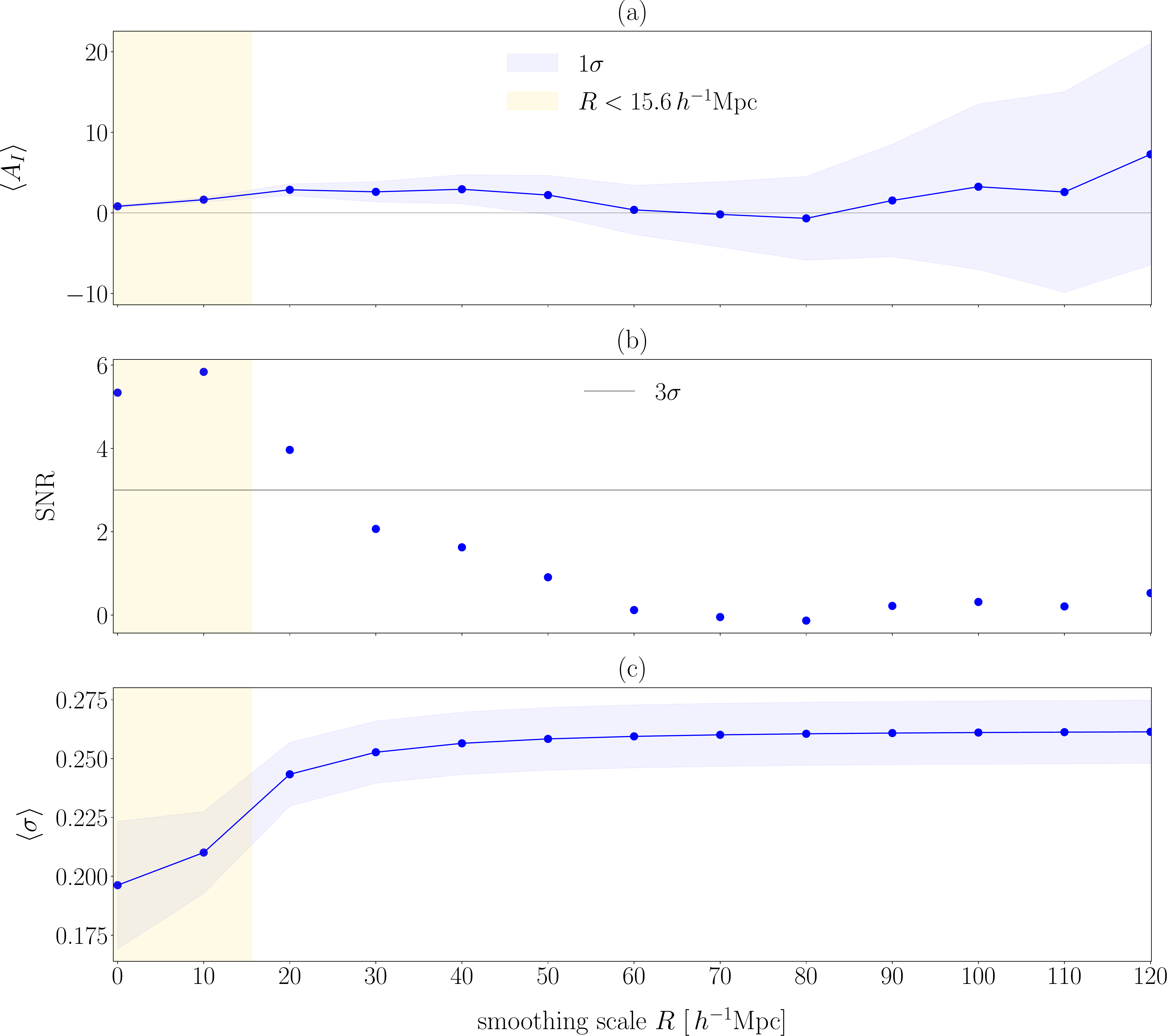}
    \caption{(a) The linear alignment amplitude posterior mean as a function of smoothing scale. (b) The signal-to-noise ratio, $\mathrm{SNR} = \langle A_I \rangle / \sqrt{\langle A_I^2 \rangle - \langle A_I \rangle^2}$, of the linear alignment amplitude as a function of smoothing scale. The uncertainty is not visible. The horizontal line represents the $3\sigma$ threshold. (c) The jointly inferred posterior mean of the root mean square random ellipticity noise. The blue and yellow windows represent 1 standard deviation and scales smaller than the inference resolution, respectively.}
    \label{fig:A_I}
\end{figure}

The linear alignment amplitude typically reported in the literature \citep[e.g.][]{Singh:2015} is derived from $C$ as follows
\be
A_I = \frac{C}{2{\Om}'\R C'\rhocr'},
\label{eq:rescaling}
\ee
where $C'\rhocr'=0.0134$, ${\Om}'=0.282$, and $\R$ is the shear responsivity. Since this quantity is specific to the galaxy shape catalog, we adopt a shear responsivity of $\R=0.87$ following \citep{Singh:2015}. Here, the primes indicate that these parameters are assumed in another study.

The linear alignment amplitude inference as a function of smoothing scale is shown in \reffig{A_I}(a). At sharp-k smoothing $R = 20$\Mpch{} applied to the tidal fields with grid resolution 15.6\Mpch{}, we find $A_I = 2.9 \pm 0.7$ from both ellipticity components. As a test, we infer the linear alignment amplitude for each ellipticity components separately and then combine the amplitudes using inverse-variance weighting. For the linear alignment amplitude constrained with the first component, we find $A_{I,1} = 3.2 \pm 1.0$, whereas from the second component we find $A_{I,2} = 2.6 \pm 1.0$. The amplitudes are consistent within $0.4\sigma$. Considering both ellipticity components, we obtain $A_I = 2.9 \pm 0.7$, consistent with the joint inference. 

\citep{Singh:2015} reported $A_I = 4.6 \pm 0.5$ for the LOWZ sample using the primordial alignment scenario. Employing the same scenario by using the estimate of the tidal field at $z=0$ we find $A_I = 3.2 \pm 0.8$, which is consistent with $<1.5\sigma$ with \citep{Singh:2015}. Compared to \citep{Singh:2015}, the variance of our intrinsic alignment posterior consists of the intrinsic dispersion in the observed ellipticity sample and the uncertainty propagated in the tidal field realizations. The latter consists of the observational uncertainties associated with the spectroscopic galaxy sample used to constrain the \codefont{BORG} realizations. Further, compared to the configuration in \citep{Singh:2015}, we applied a lower redshift cutoff and removed galaxies on the survey mask. This treatment leaves $\sim10\%$ fewer galaxies in our sample, which is equivalent to 5\% contribution to Poisson uncertainty. Factors that may drive this difference can further stem from the fact that we jointly infer the random shape noise and report our result at a single scale. Finally, the assumptions that enter our likelihood and the non-Gaussian features in the tidal field that we account for through our field-level inference likely have an impact. Finally, the cosmological parameters assumed here and in \citep{Singh:2015} are different. As a result, we expect a discrepancy between \citep{Singh:2015} and our result, though still within error bars.

On sufficiently large scales, the linear alignment strength is expected to be independent of scale. As described in \refapp{smoothing_correction}, we expect that solely the effect of filtering introduces scale dependence in our measurement, if unaccounted for. \reffig{A_I}(a) shows that our results for $A_I$ are indeed consistent with being $R$-independent for $R\gtrsim 20$\Mpch{}. On smaller scales, we expect a dependence on the filter scale $R$. First, because of nonlinear and scale-dependent (higher-derivative) corrections to the linear alignment model. Second, because the fundamental resolution of the forward model employed in the density field inference is $15.6$\Mpch{} and therefore smaller-scale processes are not modeled. We therefore expect that scales smaller than the fundamental resolution of the tidal fields will be biased by the aforementioned effects. On large scales, even though the linear alignment amplitudes at $R=60-120$\Mpch{} are consistent with a constant amplitude, fluctuations are visible. These are likely due to the fact that relatively few modes remain after these very large-scale sharp-k filters. These in turn are coupled by the survey window, which makes the Fourier modes we probe highly non-uniform, leading to the mildly correlated features seen in $A_I$ as a function of $R$ on very large scales. Therefore, our results vary, albeit non-significantly, as a function of scale, first because of the survey mask on large scales and second, because of unmodeled physics on scales smaller than the fundamental resolution of our tidal fields. Further, 65\% of the grid elements populated by galaxies contain only one galaxy. This suggests that our assumption of independent shape noise among galaxies is a good approximation. The rest of the grid elements are populated by up to 4 galaxies, with decreasing rate as a function of galaxy density. Given that the galaxy shape noise in these grid elements is correlated, we expect a lower amplitude than the one we report here.

The signal-to-noise ratio (SNR) for the intrinsic alignment amplitude is shown in \reffig{A_I}(b). It is defined as $\mathrm{SNR} = \langle A_I \rangle / \sqrt{\langle A_I^2 \rangle - \langle A_I \rangle^2}$. The detection of linear alignment is $4\sigma$ at $R=20$\Mpch{}. The ellipticity variance posterior mean, jointly inferred with the linear alignment amplitude is shown in \reffig{A_I}(c). We obtain $\sigma = 0.24 \pm 0.01$ at $R=20$\Mpch{}, constrained with both ellipticity components. Our results agree with \citep{2021MNRAS.504.4312G} at $0.2\sigma$. As discussed in \refsec{data_model} and seen from \refeq{datamodel}, our ellipticity variance only includes a contribution from the random noise component $\eps^{\rnd}$. Let us assume that the distribution of $\eps^{\rnd}$ can be approximated by the standard deviation of the observed ellipticities $e^{\obs}$. This fit then yields a $\sigma_{\rnd} = 0.26$, consistent with the value we infer. In order to obtain this fit, we consider both ellipticity components separately and then average the resulting standard deviations. Note that this test is separate from our analysis and is used only as a sanity check. In our analysis we jointly infer the random ellipticity component along with the intrinsic alignment amplitude, in order to allow for a broader posterior in case outliers are present. We note that the mean ellipticity measurement standard deviation is $\langle \sigma_\mathrm{msm} \rangle = 0.008\ll\sigma^\mathrm{rnd}$. Our inferred variance is consistent with the presumption that the noise in our cross-correlation is dominated by random processes, which are universal across the galaxy sample.

\begin{figure}
    \centering
    \includegraphics[width=\textwidth]{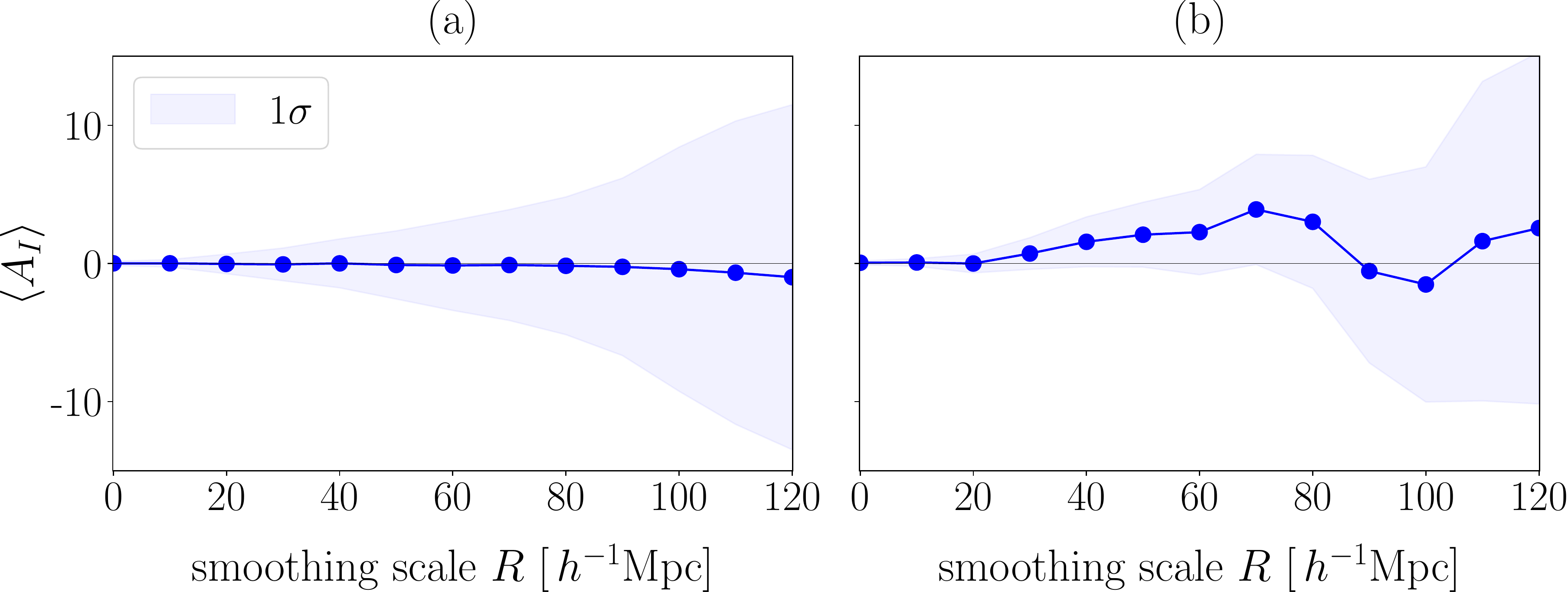}
    \caption{(a) Null test on the linear alignment amplitude performed by randomly rotating galaxy ellipticities, considering 100 datasets. (b) Null test on the linear alignment amplitude performed by randomly rotating galaxy ellipticities, considering 1 dataset. The blue window represents one standard deviation on the linear alignment amplitude posterior.}
    \label{fig:A_I_null}
\end{figure}

We have also performed a null test using randomly-rotated galaxy shapes for every galaxy. The null test indicates to what extent a spurious signal can arise even in lack of correlation between galaxy shapes and the tidal field; for this purpose, considering a single ellipticity component is sufficient. We use 10 tidal field realizations and 100 randomly-rotated galaxy shapes per galaxy. Our results are shown in \reffig{A_I_null}(a). They show that  the signal is consistent with a null detection in absence of any correlation between the galaxy shapes and the underlying tidal field. This suggests that our signal on scales of 20\Mpch{} ($1.4\degree$ at the median redshift of the ellipticity sample) constitutes a significant detection. In \reffig{A_I_null}(b), we show the same test considering one set of randomly-rotated ellipticities. The noticeable fluctuations are due to dataset variance.

\begin{figure}
    \centering
    \includegraphics[width=0.96\textwidth]{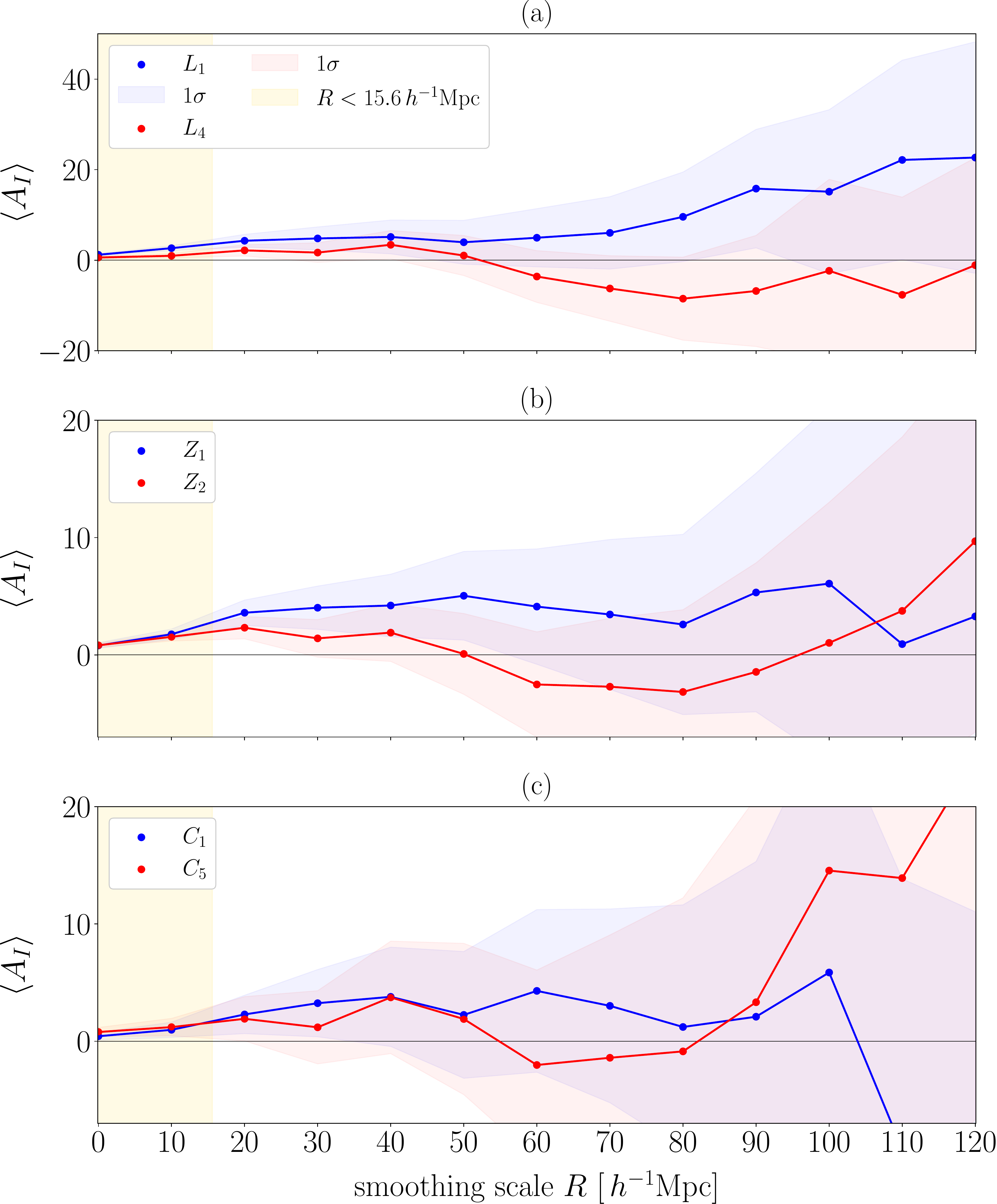}
    \caption{(a) The linear alignment amplitude as a function of smoothing scale for the brightest ($L_1$) and faintest ($L_4$) galaxy sub-sample. (b) The linear alignment amplitude as a function of smoothing scale for two redshift sub-samples. $Z_1$ covers the range $0.21 < z < 0.29$ and $Z_2$ the range $0.29 < z < 0.36$. (c) The linear alignment amplitude as a function of the smoothing scale for the bluest ($C_1$) and reddest ($C_5$) sub-sample.}
    \label{fig:L_cuts}
\end{figure}

We then examine the evolution of the linear alignment amplitude with luminosity, which previous studies have detected within luminous red galaxies \citep{Singh:2015,Joachimi:2011,Hirata:2007}. In search of such a trend, we split our LOWZ sample into four luminosity bins, based on the $r$-band absolute magnitude. Following the percentile cuts reported previously \citep[Table 2,][]{Singh:2015}, the $L_1$ sub-sample contains the brightest 20\% of galaxies and $L_4$ contains the faintest 60\%. Both samples consist of approximately the same number of galaxies. The results for the instantaneous alignment scenario are shown in \reffig{L_cuts}(a). We find a mild evolution of the linear alignment amplitude with luminosity, as the amplitude does not differ more than $3\sigma$ between any luminosity sub-sample at any scale. Using the primordial alignment scenario for comparison, the intrinsic alignment amplitude per sub-sample is consistent with previous findings \citep{Singh:2015}. Our results are $\sim 2\sigma$ away from the results reported in \citep[Table 2,][]{Singh:2015} for the $L_1$ sub-sample and within $\sim 0.1 \sigma$ for the $L_4$ sample. We attribute the discrepancy in the $L_1$ sample to the fact that we consider scales $\geq 20$ \Mpch{} in our estimate, whereas \citep{Singh:2015} does so on scales $\geq 6$ \Mpch{}, where strong nonlinearities are present. We report a $2.3\sigma$ trend of brighter galaxies toward stronger alignment amplitudes at $R\gtrsim 20$\Mpch{}. 

We proceed with the evolution of the linear alignment amplitude with redshift. To this end, we split the LOWZ sample into two redshift samples in the ranges z = 0.21 -- 0.29 and z = 0.29 -- 0.36. The results for the instantaneous alignment scenario are shown in \reffig{L_cuts}(b). We find no significant redshift evolution of the linear alignment amplitude with redshift. However, as we probe a limited redshift range, the intrinsic alignment trend as a function of redshift carries little significance.

In order to study the evolution of the intrinsic alignment amplitude with color, we divide the LOWZ galaxies into five sub-samples, based on their $g-i$ color index. The sub-samples are defined according to earlier work \citep[Table 2,][]{Singh:2015}, where $C_1$ is the bluest and $C_5$ the reddest sample. The results for the instantaneous alignment scenario are shown in \reffig{L_cuts}(c). We observe no significant color dependence of the linear alignment amplitude. Switching to the primordial alignment scenario for easier comparison, our findings are consistent with past results \citep{Singh:2015}. Given that the sample consists of luminous red galaxies, a lack of color evolution is expected. 

\section{Discussion and conclusions}\label{sec:discussion}

We have constrained the linear alignment model of elliptical galaxies by correlating their observed ellipticities as reported in the SDSS-III BOSS and LOWZ sample and the tidal fields constrained by the \codefont{BORG} algorithm with SDSS-III BOSS spectroscopic galaxy data. At smoothing scale $R=20$\Mpch{}, we have inferred a linear alignment amplitude of $A_I = 2.9 \pm 0.7$, at $4\sigma$ significance level. On scales $R \lesssim$ 20\Mpch{}, the linear alignment amplitude is expected to be biased due to contributions from both nonlinear and unresolved processes. We report a mild evolution of the linear alignment amplitude with luminosity and no evolution with redshift or color within this luminous red galaxy sample. The mean ellipticity variance, which we have assumed to be universal for all galaxies and to be dominated by random processes, is $\sigma = 0.24 \pm 0.01$ at $R=20$\Mpch{}.
Within our modeling framework, we expect that deviations from the true linear alignment amplitude occur due to systematic effects at small scales. First, as the inference grid resolution is 15.6\Mpch{} and we consider a gravity-only forward model, some nonlinear features are not captured in the original \codefont{BORG} inference \citep{Lavaux:2019}. Further, unmodeled astrophysical processes during galaxy formation may affect the inference on scales approaching the grid scale \citep[e.g.][]{2018MNRAS.480.3962C,2019JCAP...03..020S,2021MNRAS.503.3796D}. Second, our linear alignment assumption breaks down on scales where the large-scale structure becomes nonlinear. In this case, galaxy shapes are expected to become a nonlinear function of the tidal field. On small scales, the Gaussian ellipticity likelihood and the assumption of independent shape noise will likewise begin to break down.

In order to estimate the large-scale alignment strength, and prevent leakage of information from smaller scales, we employ a sharp-k filter.
Smoothing scales of $R\geq20$ \Mpch{} are sufficiently larger than the typical scale of strong nonlinearities in large-scale structure. We therefore do not expect that our results at those scales are significantly affected by the effects mentioned above. This is supported by the apparent scale-independence of our inferred alignment amplitude at $R \geq 20$\Mpch.

Going forward, the field-level approach that we have presented here allows for improvement on different aspects of our modeling. First, our approach -- based on a field-level, forward-modeling framework like the \codefont{BORG} algorithm -- facilitates the incorporation of nonlinear corrections to both gravitational evolution and the intrinsic alignment model \citep[for the latter see e.g.][]{2019PhRvD.100j3506B,2021MNRAS.501.2983F}. The importance of this prospect can be seen in \reffig{A_I}b: the information content of the intrinsic alignment posterior increases at smaller scales, where nonlinearities in intrinsic alignment and the tidal fields become important \citep[e.g.][]{2008ApJ...681..798L}. In particular, if the tidal field was Gaussian, our framework would be expected to yield similar results to 2-point correlation approaches. However, as non-Gaussianities in the matter density distribution become significant, the higher-order statistics that we incorporate through the \codefont{BORG} algorithm can provide a wealth of additional information on small scales.

Further, our approach allows for a joint inference of intrinsic alignment and weak lensing \citep[e.g.][]{2021MNRAS.502.3035P, 2021arXiv210514699F}, which is particularly useful at high redshifts \citep[e.g.][]{2021arXiv210914297T}. Owing to the full treatment of known and unknown systematic effects and observational uncertainties within \codefont{BORG}, this aspect guarantees a fully self-consistent inference of both intrinsic alignment and weak lensing. Systematic effects were treated by using a likelihood that marginalizes over foreground contaminations \citep{2019A&A...624A.115P} and employing the technique of template matching \citep{2017A&A...606A..37J}. Observational uncertainties stemming from the survey geometry and luminosity function are directly modeled and accounted for within \codefont{BORG} through the data likelihood \citep[Fig. 1,][]{Lavaux:2019} and the resulting uncertainty is propagated to the density field inference. In our study, we marginalize over the tidal field realizations that vary due to the aforementioned uncertainties. As a result, all these uncertainties are propagated into the intrinsic alignment inference.

Note that our intrinsic alignment amplitude posterior considers the tidal field on scales $\geq R$, $R$ being the smoothing scale of the tidal fields. Therefore, information from scales $> R$ is already contained in those with smaller smoothing scales. In order to extend our analysis to smaller scales, one would choose an intrinsic alignment model that accurately captures nonlinearities and increase the resolution of the \codefont{BORG} inference.

Our approach facilitates the use of intrinsic alignment as a cosmological probe, e.g. in conjunction with redshift-space distortions, Baryon Acoustic Oscillations and primordial non-Gaussianity \citep[e.g.][]{Chisari:2013,2019A&A...621A..69R,2022arXiv220308838A}. Cosmological constraints are expected to improve when smaller scales are included in the analysis \citep[e.g.][]{2016PhRvD..94l3507C}. On these scales, galaxy bias, which affects the intrinsic alignment signal, is expected to be scale-dependent. Within \codefont{BORG}, scale-dependent galaxy bias can be inferred jointly with the large-scale structure (and consequently, the tidal shear) conditional on galaxy observations \citep{2022arXiv220308838A}. Moreover, 2-point analyses consider the matter power spectrum at a given effective redshift within a galaxy sample. Field-level approaches can consider the tidal field at the redshift of each individual galaxy, if the evolution of intrinsic alignment with redshift is known. Further, cross-correlations of shape measurements with \codefont{BORG} inferences -- where $\Lambda$CDM and a certain structure formation model are assumed -- can be used as a test of these models, provided that the assumed intrinsic alignment and gravity models are satisfactory descriptions of the physics on the relevant scales. Finally, the cross-correlation between the tidal field and galaxy shape we infer is local in real space. This renders the isolation and testing of map-level systematic effects easier.

\section*{Data and software availability}

For data requests, please contact the Aquila Consortium (\url{https://www.aquila-consortium.org}). Software products may be made available by E.T. and N.-M.N upon reasonable request. 

\section*{Acknowledgements}

We thank the Reviewer for their careful review, which helped improve the manuscript. The authors thank Rachel Mandelbaum and Sukhdeep Singh for kindly providing the SDSS-III BOSS/LOWZ shape measurements catalog. We further thank Adam Andrews, Deaglan Bartlett, Supranta Sarma Boruah, Alan Heavens, Dragan Huterer, Eiichiro Komatsu and Natalia Porqueres for providing feedback on the manuscript. This research utilized the HPC facility supported by the Technical Division at the Department of Physics, Stockholm University and the Max Planck Computing and Data Facility. This work was enabled by the research project grant ‘Understanding the Dynamic Universe’ funded by the Knut and Alice Wallenberg Foundation under Dnr KAW 2018.0067. JJ acknowledges support by the Swedish Research Council (VR) under the project 2020-05143 -- "Deciphering the Dynamics of Cosmic Structure". MN and FS acknowledge support from the Starting Grant (ERC-2015-STG 678652) \enquote{GrInflaGal} of the European Research Council. This work has made use of the Infinity Cluster hosted by Institut d'Astrophysique de Paris.
This work is conducted within the Aquila Consortium (\url{https://aquila-consortium.org}). 

\appendix

\section{Smoothing correction of the linear alignment amplitude}\label{app:smoothing_correction}

In our inference of the alignment amplitude, we introduce a sharp-k filter in order to remove small scales that are affected by nonlinearities. Naive comparison between ellipticities and tidal fields to determine the linear alignment amplitude would appear as scale dependent. This phenomenon is due to the presence of the NGP and CIC implicit filtering, whose impact on either ellipticities or on our tidal amplitudes would be neglected. 

We now propose a model and a correction for this scale dependence. To simplify the discussion, we assume that galaxy ellipticities are represented by a continuous field. A more detailed description in the presence of selection effects is postponed to future work. As galaxies allow us to observe ellipticities only over their actual scale, we consider that observed galaxy ellipticities are filtered with a top-hat (TH) kernel of size $R_\mathrm{TH}$. The assignment of galaxies to the grid is made with an NGP kernel
\be
W_\mathrm{NGP}(\mathbf{x},R) = \prod_{i=1}^3 \begin{cases} 1 & \text{if } |x_i| < R/2 \\
0 & \text{otherwise,}      
\end{cases}
\ee

where $\{ x_i : i \in \{1,2,3\}\}$ are comoving Cartesian coordinates. The \codefont{BORG} inference and resimulation by construction employ a cubic CIC filter with the common resolution of 15.6\Mpch for the side of each cubic element. The CIC filter is defined as
\be
W_\mathrm{CIC}(\mathbf{x},R) =  \prod_{i=1}^3 \begin{cases} 
1-|x_i|/R & \text{if } |x_i| < R \\
0 & \text{otherwise.}      
\end{cases}
\ee
Finally, we apply a sharp-k filter on the resulting tidal field. This operation leads to removing modes with $k> 1/R$.

Here, we will denote by $[Q]_{\rm X}$ a quantity $Q$ that is convolved  with a kernel $X$ of a given size and by $W_{\rm X}$ the corresponding convolution function. We can then write the data model used in the inference as
\be
{[\mathbf{e}^\obs]}_{\mathrm{TH}} = \frac{C}{4\pi G}\left[T_\pm\right]_{\mathrm{CIC-SK-NGP}} + [\epsilon^{\mathrm{rnd}}]_{\mathrm{TH}},
\label{eq:smoothing_1}
\ee
where $C$ indicates the linear alignment amplitude which refers to the filtered tidal field. $\epsilon^{\mathrm{rnd}}$ is a field that contains all the parts that are not modeled through the tidal field. We assume that its values are Gaussian-distributed and on average decorrelated from the tidal fields $T_\pm$. However, it is still a continuous field. Suppose that the linear alignment model holds on all scales, i.e. galaxy shapes are linearly proportional to the tidal field at their location. The data model that associates the two fields reads
\be
{[\mathbf{e}^\obs]}_{\mathrm{TH}} = \frac{C^{\rm corr}}{4\pi G}\left[T_\pm\right]_{\mathrm{TH}} + [\tilde\epsilon^{\mathrm{rnd}}]_{\mathrm{TH}},
\label{eq:smoothing_1b}
\ee
where now we have introduced the corrected linear alignment amplitude $C^{\rm corr}$, and we have indicated the different shape noise field with a tilde. 
\begin{figure}[t]
    \centering
    \includegraphics[width=\textwidth]{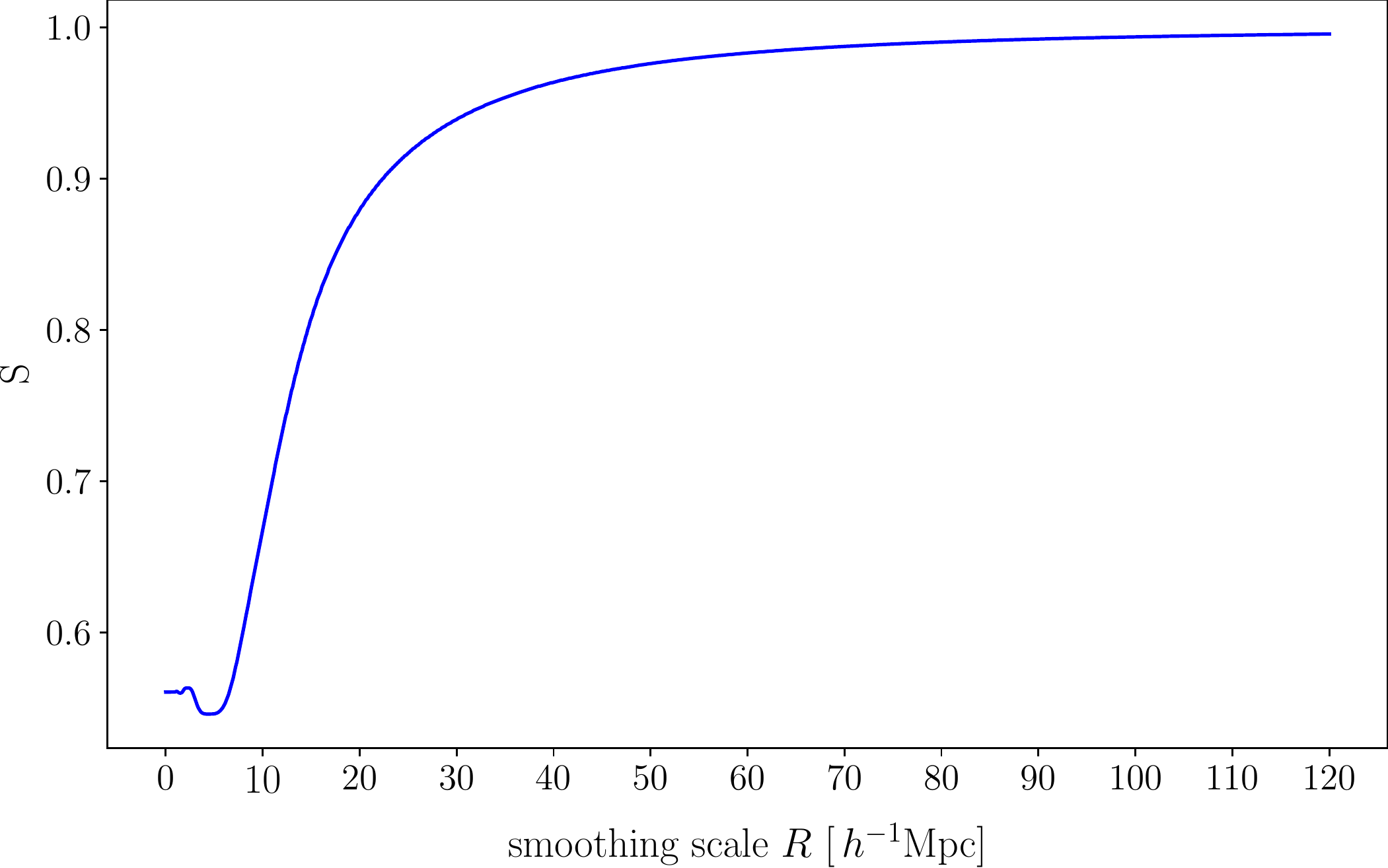}
    \caption{The smoothing correction of \refeq{smoothing}, as a function of the sharp-k smoothing scale.}
    \label{fig:S_factor}
\end{figure}
Our goal is to derive $C^{\rm corr}$ in terms of $C$. For this, we multiply both sides of \refeq{smoothing_1b} by $[T_\pm]_\mathrm{CIC-SK-NGP}$ and take the ensemble average. Note that the random noise component changes from $\epsilon^{\mathrm{rnd}}$ in \refeq{smoothing_1} to $\tilde\epsilon^{\mathrm{rnd}}$ in \refeq{smoothing_1b}, because it now refers to the \textit{corrected} intrinsic alignment amplitude. This suggests that at this step, the artificial scale dependence we introduced when smoothing, is now accounted for. Since the shape measurement and the random noise are both uncorrelated with the smoothed tidal field, we obtain
\be
\langle [T_\pm]_\mathrm{CIC-SK-NGP} [\mathbf{e}^\obs]_\mathrm{TH} \rangle = \frac{C^\mathrm{corr}}{4\pi G} \langle [T_\pm]_\mathrm{CIC-SK-NGP} [T_\pm]_\mathrm{TH} \rangle.
\label{eq:smoothing_2}
\ee
An analogous relation involving $C$ is obtained from \refeq{smoothing_1}. 
Assuming that the dynamics is provided by the linear perturbation theory and that there are no mask or selection effects, the bracketed terms can be evaluated as integrals over the linear matter power spectrum, $P_L(\mathbf{k})$, defined as
\be
\langle \delta(\vk) \delta(\vk')\rangle = (2\pi)^3 \delta_D(\vk+\vk') P_L(k)\;.\label{eq:linear_theory}
\ee
This assumption is accurate in the limit of large smoothing, since this filter cuts off high-$k$ modes. Therefore, we model the dependence on the filtering scale of the linear alignment amplitude by the linear relation
\be
C^{\rm corr}(k) = S(k)\, C(k),
\label{eq:smoothing_correction}
\ee
where
\be
S(k) = \frac{\langle [T_\pm]_\mathrm{CIC-SK-NGP}(\vk) [T_\pm]_\mathrm{CIC-SK-NGP}(\vk) \rangle}{\langle  [T_\pm]_\mathrm{CIC-SK-NGP}(\vk) [T_\pm]_\mathrm{TH}(\vk)\rangle}\;.
\label{eq:filtering_scale_factor}
\ee

Using \refeq{linear_theory} and \refeq{filtering_scale_factor}, we give the explicit expression for $S(k)$
\be
S(k) = \frac{\int \mathrm{d}^3\mathbf{k}\; P_L(\mathbf{k})W^2_\mathrm{CIC}(\mathbf{k}, R_\mathrm{CIC})W^2_\mathrm{SK}(\mathbf{k}, R)W^2_\mathrm{NGP}(\mathbf{k}, R_\mathrm{CIC})}{\int \mathrm{d}^3\mathbf{k}\; P_L(\mathbf{k})W_\mathrm{CIC}(\mathbf{k}, R_\mathrm{CIC})W_\mathrm{SK}(\mathbf{k}, R)W_\mathrm{NGP}(\mathbf{k}, R_\mathrm{CIC})W_\mathrm{TH}(\mathbf{k}, R_\mathrm{TH})}.
\label{eq:smoothing}
\ee
We note that $S \to 1$ when $R \gg R_{\rm CIC}, R_{\rm TH}$. We approximate the cubic CIC filter at the grid resolution, 15.6\Mpch{}, with a spherically-symmetric CIC kernel of radius 9.7\Mpch{}. The TH kernel radius, representing the observed ellipticity smoothing, is taken to be $R_\mathrm{TH}=1$\Mpch{}. However, the precise value of this scale can be ignored as long as it is much smaller than $R_{\rm CIC}$. The smoothing correction as a function of the SK smoothing scale, $R$, is shown in \reffig{S_factor}. The power spectra are considered at the median redshift of the galaxy sample, $z=0.26$. As expected, the larger $R$ is, the smaller the correction is (i.e. the closer $S$ becomes to 1). 

\section{Posterior derivation}\label{app:posterior_formulation}

In this appendix, we provide the derivation of the marginalized posterior for the linear alignment amplitude, as described in \refsec{LA_model}. In what follows, $N_G^3=256^3$ is the number of grid elements in the \codefont{BORG} inference. $\mathbf{T}$ denotes the 6 independent tidal field components at each grid element. We start from the joint marginal posterior of the intrinsic alignment amplitude and the unknown random noise component
\be
\P\left(C,\sigma^2\Big|e\right)  = \int \mathrm{d}^{N_G^3}\mathbf{T}\; \P\left(C,\, \sigma^2, \mathbf{T} \Big|e \right)\; 
\label{eq:posterior_marginalisation}
\ee
which we recast into
\be
\P\left(C,\sigma^2\Big|e\right)  \propto \int \mathrm{d}^{N_G^3}\mathbf{T}\; \P\left(e \Big|C,\sigma^2,\, \mathbf{T} \right)\, \P\left(C\right) P(\sigma^2)\P(\mathbf{T})\;.
\label{eq:posterior_singlesample}
\ee
The above formulation is based on the assumption that the priors on the linear alignment amplitude, the tidal field and the random noise are independent. The \codefont{BORG} inference approximates $\P\left(\mathbf{T}\right)$ by an ensemble of $N_s$ MCMC samples
\be
\P\left(\mathbf{T} \right) \approx \frac{1}{N_s} \sum_{s=1}^{N_s} \delta_{\rm D}^{N_G^3}\left(\mathbf{T}-\mathbf{T}_{s}^\codefont{BORG}\right), \,
\ee
where $\delta_{\rm D}^{n}$ denotes the $n$-dimensional Dirac delta distribution. The above equation indicates that we consider each sample in the tidal field posterior to be given by each \codefont{BORG} tidal field. We then write the joint marginal posterior as
\bab
\P\left(C,\sigma^2\Big|e\right)  \propto \sum_{s=1}^{N_s} \P\left(e \Big|C,\sigma^2,\, \mathbf{T}_{s}^\codefont{BORG} \right)\, \P\left(C\right) P(\sigma^2)\P(\mathbf{T}_{s}^\codefont{BORG}).
\label{eq:joint_marginal_posterior}
\eab
We then assume a Gaussian likelihood for observed ellipticities. The likelihood allows us to consider the tidal field only at those grid cells that are populated with galaxies. To simplify notation, we designate the tidal field values at the location of galaxies per realization as $T_s$. Note that the two ellipticity components are modeled independently and are Gaussian-distributed. We rewrite the Gaussian likelihood for a single realization as \citep{Nguyen:2020a}
\be
\ln\P\left(e|C,\, \sigma^2,T_s \right) = -\frac{\gamma_{s}}{2} + \frac{\mu_{s}^2}{2\,\nu_{s}^2} - \frac{\delta_{s}}{2} - \frac{\left(C-\mu_{s}\right)^2}{2\,\nu_{s}^2},
\label{eq:LA_ln_likelihood_short}
\ee
with
\be
\gamma_{s} = \sum_{g=1}^{2N_g} \frac{e_{g}^2}{\sigma_g^2}
\quad\text{,}\quad 
\delta_{s} = \sum_{g=1}^{2N_g} \ln\sigma_g^2
\label{eq:gamma_delta}
\ee
Note that the sums run to $2N_g$, as we have now considered the product likelihood of \refeq{likelihood_singlesample}, in which we consider both components of observed and projected ellipticities per galaxy. 

Given that the first and third terms on the RHS of \refeq{LA_ln_likelihood_short} are independent of the linear alignment amplitude, we ignore them and recast the single-realization posterior for the linear alignment amplitude into the following
\be
\P\left(C, \sigma^2 \Big|e, T_s\right) \propto   \P(C)  \frac{\lambda_{s}}{\sqrt{2\pi\,\nu_{s}^2 }}\exp\left[-\frac{\left(C-\mu_{s}\right)^2}{2\nu_{s}^2}\right].
\label{eq:amplitude_posterior_1}
\ee

Marginalizing over the tidal field realizations, we arrive at \refeq{amplitude_posterior_2}. The linear alignment amplitude posterior mean can be analytically derived for a uniform prior on $C$ as
\be
\<C\> = \sum_{s=1}^{N_s} \lambda_{s} \mu_{s}.
\label{eq:MAP_mean}
\ee
Finally, the analytical expression for the variance of the posterior mean, ${\sigma^2_C}$, reads
\be
{\sigma^2_C} = \sum_{s=1}^{N_s} \lambda_{s}  \left[\nu_{s}^2  +  \, \left(\mu_{s} - \<C\> \right)^2\right]
\label{eq:MAP_variance}
\ee

\section{Validation tests}\label{app:validation}

In this section we describe the validation tests of the method presented in \refsec{posterior}. We test our sampling algorithm using mock data and a comparison to the analytical posterior moments presented in Appendix \ref{app:posterior_formulation} for fixed variance. We test the convergence using a Gelman-Rubin test. We show that the linear alignment amplitude posterior mean and variance are insensitive to the choice of variance prior. Finally, we present a posterior predictive check using one realization of the tidal field.

\subsection{Mock data test}\label{app:mock_data}

\begin{figure}
    \centering
    \includegraphics[width=\textwidth]{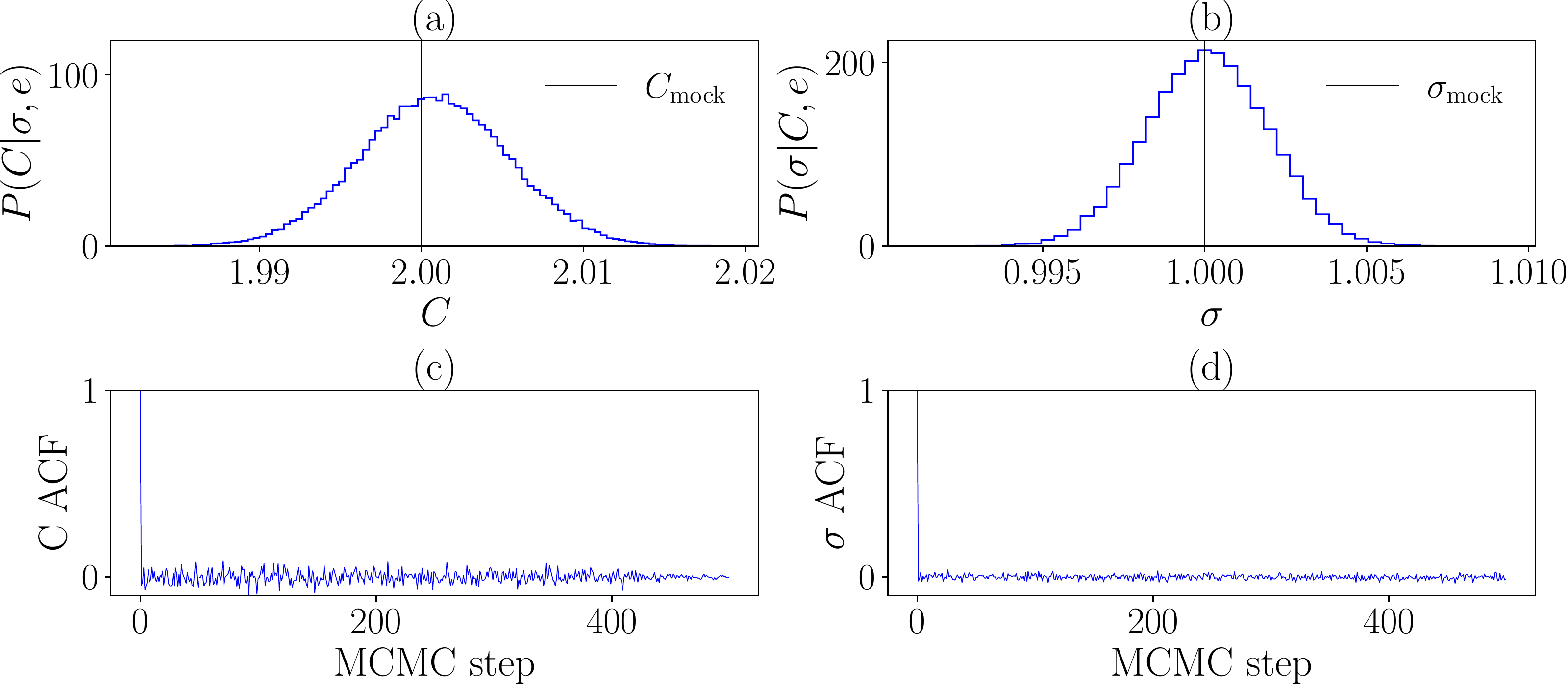}
    \caption{(a) The linear alignment amplitude posterior from mock data generated from a Gaussian with known linear alignment amplitude, $C_\mathrm{mock} = 2$ and unit variance, considering only 1 realization of the tidal field. (b) The universal variance posterior for the same mock data. The ground truth is $\sigma^2_\mathrm{mock} = 1$. (c) The autocorrelation function of the mock linear alignment amplitude MCMC chain. (d) The autocorrelation function of the mock variance MCMC chain.}
    \label{fig:mock_tests}
\end{figure}

\begin{figure*}
    \centering
    \includegraphics[width=\textwidth]{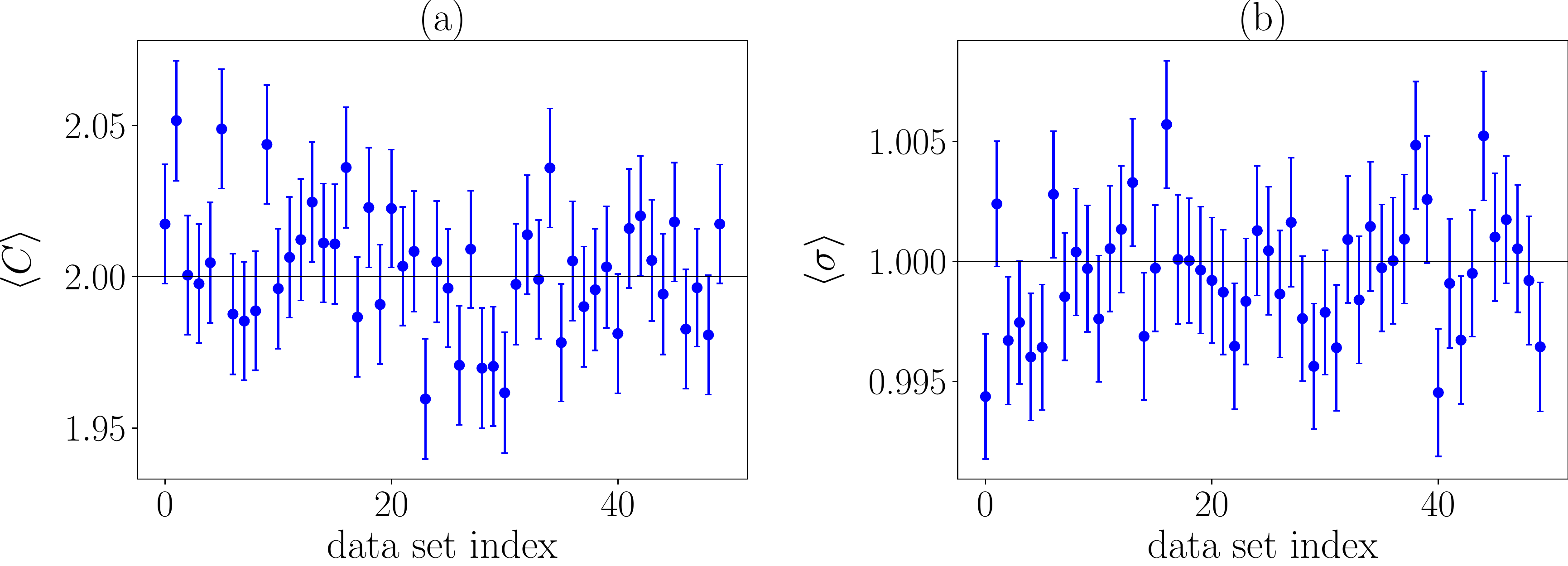}
    \caption{Fluctuations of the (a) linear alignment amplitude and (b) variance posterior means as a function of mock dataset index.}
    \label{fig:dataset_variance}
\end{figure*}

We generate a sample of mock ellipticities of the same size as the observed ellipticities sample. We consider 5 realizations of random tidal fields. The mock ellipticities are drawn from a Gaussian distribution with known input amplitude $C_\mathrm{mock}$ and unit variance, $N(e|C_\mathrm{mock},T_s, 1)$. The prior on variance is $\mathrm{IG}(10^{-3}, 10^{-3})$. The correction factor for mock data is equal to unity on all scales. Each chain consists of 10~000 steps. The linear alignment amplitude posterior is shown in \reffig{mock_tests}(a) and the corresponding autocorrelation function (ACF) is shown in \reffig{mock_tests}(c). The variance posterior is shown in \reffig{mock_tests}(b) and the corresponding autocorrelation function is shown in \reffig{mock_tests}(d). As can be seen from \reffig{mock_tests}, there is an offset between the posterior means and the corresponding mock values. We test whether the offset is due to mock dataset variance, by generating 50 mock datasets with differing seeds in chains of 5~000 steps and monitoring the average behavior of the posterior means around the mock linear alignment amplitude and variance. Our results are shown in \reffig{dataset_variance}. The average offset between the input linear alignment amplitude and the inferred mean averaged across the datasets is $\sim10^{-5}$. The corresponding value for the variance posterior is $\sim10^{-3}$. We therefore conclude that the offset is due to dataset variance.

\subsection{Gelman-Rubin test}

In order to test the convergence of the MCMC chains, we perform a Gelman-Rubin test \citep{1992StaSc...7..457G}. The results were obtained for 10 independent chains in which the starting point was $C = [1, 2, ..., M = 10]$, considering only one ellipticity component. We performed the tests on a mock dataset of the same sample size as the real data, but with unit variance. The variance prior was the IG(1, 0.1). Every chain ran for 10~000 steps, corresponding to $2N$ in the notation of this test. We computed the potential scale reduction factor (PSRF)
\be
    R = \sqrt\frac{V}{W},
\ee
where
\be
    V = \frac{N-1}{N}W + \frac{M+1}{MN}B,
    \quad\text{,}\quad 
    B = \frac{N}{M-1}W \sum_{m=1}^M (\hat{a}_m - \hat{a})^2,
\ee
where $\hat{a}_m$ are the posterior means per chain, $\hat{a}$ is the posterior mean averaged over the chains and $W$ is the posterior variance averaged over the chains. A PSRF close to unity is indicative of convergence. Sampling variability can be accounted for by recasting the PSRF into $R_c = \sqrt{(d+3)/(d+1)} R$. As $d$ represents the degrees of freedom estimate of a t distribution, in our case it is given by $d = N_g - 1$. As a result, the variance sampling correction is close to unity. If $R_c<1.2$, convergence has been achieved. Our results are shown in \reftab{GR} and indicate that our sampler converges to the target distribution.

\begin{table}[]
\caption{The Gelman-Rubin test PSRF on mock and real data for one and multiple realizations. The results indicate that convergence has been reached.}
\centering
\begin{tabular}{|c|c|c|}
\hline
Data & $N_s$ & $|1-R_c|$         \\ \hline
Mock & 1    & $10^{-5}$ \\ \hline
Mock & 5    & $10^{-4}$ \\ \hline
Real & 1    & $10^{-4}$ \\ \hline
Real & 5    & $10^{-4}$ \\ \hline
\end{tabular}
\label{tab:GR}
\end{table}

\subsection{Posterior moments}

To the limit of fixed variance, the linear alignment amplitude posterior mean and variance are given by \refeq{MAP_mean} and \refeq{MAP_variance}, respectively. In the context of our implementation, where each tidal field sample is considered independently, i.e. $\lambda_{s} = 1$, the above equations reduce to the Blackwell-Rao estimators. It should be noted that even though $\lambda_{s} = 1$, the Gaussian mixture posterior components receive their corresponding weights through proper normalization. Using the configuration described in Appendix \ref{app:mock_data} for mock data generation and sampling linear alignment amplitudes with a fixed variance set to the mock data variance, we recover $C = 2.00 \pm 0.05$, which is the result from the Blackwell-Rao estimators. This indicates that the linear alignment amplitude inference is consistent with the theoretical expectation for the posterior moments.

\subsection{Sensitivity to prior}\label{subs:prior_sensitivity}

\begin{figure}
    \centering
    \includegraphics[width=\textwidth]{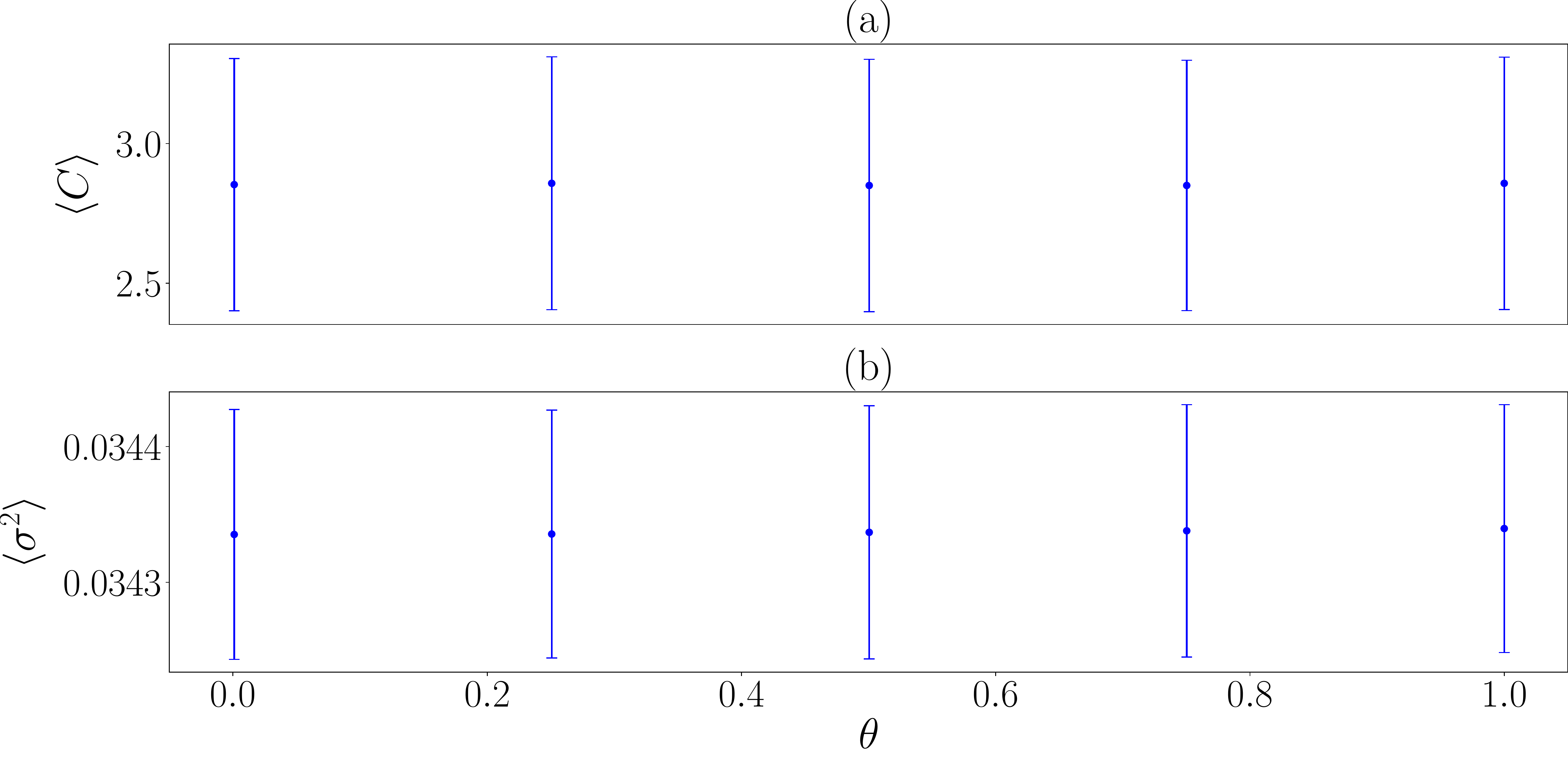}
    \caption{(a) Linear alignment amplitude and (b) variance due to random noise posterior mean as a function of an IG($\theta, \theta$) variance prior.}
    \label{fig:prior_sensitivity}
\end{figure}

In the case of Jeffrey's prior, $(\theta_1,\theta_2) = (0,0)$. However, due to the improper nature of the prior, the variance posterior also becomes improper. For this reason, we examine the sensitivity of our results to the choice of prior and find that our inference is not sensitive to different IG($\theta_1,\theta_2$) priors for $(\theta_1,\theta_2)\in(10^{-3},..,1)$, as all galaxies constrain one universal variance. Therefore, we would like to use another conjugate prior of the $\mathrm{IG}(\theta,\theta)$ kind, aiming for $\theta \rightarrow 0$, such that the prior entropy is high, indicating a prior that is as weakly informative as possible. Given the asymptotic behavior of the entropy of IG distributions at $\theta \rightarrow 0$, we explore the sensitivity of the linear alignment amplitude posterior to the choice of prior. We examine the posterior mean and standard deviation for $\theta = [10^{-3},...,1]$ on real data at 20\Mpch{}, considering one random realization. The results are shown in \reffig{prior_sensitivity} and suggest that our inference is insensitive to the prior choice in the aforementioned range.

\subsection{Posterior predictive check}\label{app:ppc}

\begin{figure*}
    \centering
    \includegraphics[width=\textwidth]{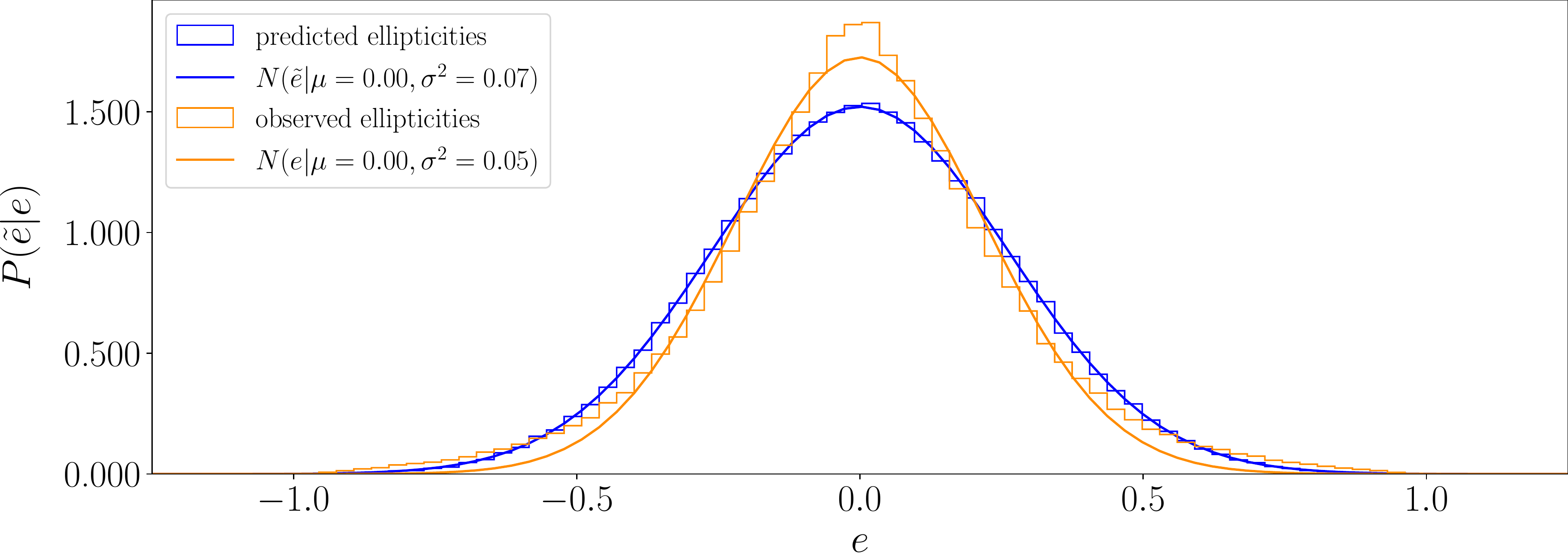}
    \caption{Posterior predictive check on the LOWZ sample using one tidal field realization at $R=$ 20\Mpch{}. The orange curves show a Gaussian fit to the inferred ellipticities.}
    \label{fig:ppc}
\end{figure*}

With this test we aim to test the suitability of the Gaussian likelihood as an approximation to the distribution of the observed ellipticities. We use one tidal field realization and run an MCMC chain for $N=10,000$ steps. We then draw $N$ ellipticities from a Gaussian distribution for each one of which we use the linear alignment amplitude and variance sample at that MCMC step as the Gaussian mean and variance, respectively. Our results for both observed ellipticities are shown in \reffig{ppc}. We perform a two-sided Kolmogorov-Smirnov (KS) test \citep[see][and references therein]{10.2307/2280095} and measure the Kullback-Leibler (KL) divergence \citep{10.1214/aoms/1177729694} of the two ellipticity samples: the observed and predicted ones by the posterior. The former yields a KS statistic of 0.03, with a p-value of 0. The latter yields a KL divergence of 0.1. Based on the p-value, we can reject the null hypothesis that the two samples are drawn from the same distribution. However, both the KS statistic of 0.03 and KL divergence of 0.1 suggest that there is little difference between the two distributions, and that the small p-value is likely driven by our large sample size.
Therefore, for the linear alignment amplitude inference, a Gaussian distribution might not be the optimal likelihood. However, our inference can be extended to different, more robust likelihoods, such as the Student's t \citep{10.2307/2331554}, in the future.

\bibliographystyle{JHEP}
\bibliography{main}

\providecommand{\href}[2]{#2}\begingroup\raggedright\begin{thebibliography}{10}

\bibitem{2002MNRAS.333..501B}
M.~L. {Brown}, A.~N. {Taylor}, N.~C. {Hambly} and S.~{Dye}, \emph{{Measurement
  of intrinsic alignments in galaxy ellipticities}},
  \href{https://doi.org/10.1046/j.1365-8711.2002.05354.x}{\emph{\mnras}
  {\bfseries 333} (July, 2002) 501--509},
  [\href{https://arxiv.org/abs/astro-ph/0009499}{{\ttfamily
  astro-ph/0009499}}].

\bibitem{Mandelbaum:2006}
R.~{Mandelbaum}, C.~M. {Hirata}, M.~{Ishak}, U.~{Seljak} and J.~{Brinkmann},
  \emph{{Detection of large-scale intrinsic ellipticity-density correlation
  from the Sloan Digital Sky Survey and implications for weak lensing
  surveys}},
  \href{https://doi.org/10.1111/j.1365-2966.2005.09946.x}{\emph{\mnras}
  {\bfseries 367} (Apr., 2006) 611--626},
  [\href{https://arxiv.org/abs/astro-ph/0509026}{{\ttfamily
  astro-ph/0509026}}].

\bibitem{Hirata:2007}
C.~M. {Hirata}, R.~{Mandelbaum}, M.~{Ishak}, U.~{Seljak}, R.~{Nichol}, K.~A.
  {Pimbblet} et~al., \emph{{Intrinsic galaxy alignments from the 2SLAQ and SDSS
  surveys: luminosity and redshift scalings and implications for weak lensing
  surveys}},
  \href{https://doi.org/10.1111/j.1365-2966.2007.12312.x}{\emph{\mnras}
  {\bfseries 381} (Nov., 2007) 1197--1218},
  [\href{https://arxiv.org/abs/astro-ph/0701671}{{\ttfamily
  astro-ph/0701671}}].

\bibitem{Joachimi:2011}
B.~{Joachimi}, R.~{Mandelbaum}, F.~B. {Abdalla} and S.~L. {Bridle},
  \emph{{Constraints on intrinsic alignment contamination of weak lensing
  surveys using the MegaZ-LRG sample}},
  \href{https://doi.org/10.1051/0004-6361/201015621}{\emph{\aap} {\bfseries
  527} (Mar., 2011) A26}, [\href{https://arxiv.org/abs/1008.3491}{{\ttfamily
  1008.3491}}].

\bibitem{Blazek:2011}
J.~{Blazek}, M.~{McQuinn} and U.~{Seljak}, \emph{{Testing the tidal alignment
  model of galaxy intrinsic alignment}},
  \href{https://doi.org/10.1088/1475-7516/2011/05/010}{\emph{\jcap} {\bfseries
  2011} (May, 2011) 010}, [\href{https://arxiv.org/abs/1101.4017}{{\ttfamily
  1101.4017}}].

\bibitem{Singh:2015}
S.~{Singh}, R.~{Mandelbaum} and S.~{More}, \emph{{Intrinsic alignments of
  SDSS-III BOSS LOWZ sample galaxies}},
  \href{https://doi.org/10.1093/mnras/stv778}{\emph{\mnras} {\bfseries 450}
  (June, 2015) 2195--2216}, [\href{https://arxiv.org/abs/1411.1755}{{\ttfamily
  1411.1755}}].

\bibitem{2022MNRAS.509.3868H}
J.~{Harnois-D{\'e}raps}, N.~{Martinet} and R.~{Reischke}, \emph{{Cosmic shear
  beyond 2-point statistics: Accounting for galaxy intrinsic alignment with
  projected tidal fields}},
  \href{https://doi.org/10.1093/mnras/stab3222}{\emph{\mnras} {\bfseries 509}
  (Jan., 2022) 3868--3888}, [\href{https://arxiv.org/abs/2107.08041}{{\ttfamily
  2107.08041}}].

\bibitem{2001MNRAS.320L...7C}
P.~{Catelan}, M.~{Kamionkowski} and R.~D. {Blandford}, \emph{{Intrinsic and
  extrinsic galaxy alignment}},
  \href{https://doi.org/10.1046/j.1365-8711.2001.04105.x}{\emph{\mnras}
  {\bfseries 320} (Jan., 2001) L7--L13},
  [\href{https://arxiv.org/abs/astro-ph/0005470}{{\ttfamily
  astro-ph/0005470}}].

\bibitem{2019PhRvD.100j3506B}
J.~A. {Blazek}, N.~{MacCrann}, M.~A. {Troxel} and X.~{Fang}, \emph{{Beyond
  linear galaxy alignments}},
  \href{https://doi.org/10.1103/PhysRevD.100.103506}{\emph{\prd} {\bfseries
  100} (Nov., 2019) 103506},
  [\href{https://arxiv.org/abs/1708.09247}{{\ttfamily 1708.09247}}].

\bibitem{2015PhR...558....1T}
M.~A. {Troxel} and M.~{Ishak}, \emph{{The intrinsic alignment of galaxies and
  its impact on weak gravitational lensing in an era of precision cosmology}},
  \href{https://doi.org/10.1016/j.physrep.2014.11.001}{\emph{\physrep}
  {\bfseries 558} (Feb., 2015) 1--59},
  [\href{https://arxiv.org/abs/1407.6990}{{\ttfamily 1407.6990}}].

\bibitem{Euclid:2013}
L.~{Amendola}, S.~{Appleby}, D.~{Bacon}, T.~{Baker}, M.~{Baldi}, N.~{Bartolo}
  et~al., \emph{{Cosmology and Fundamental Physics with the Euclid Satellite}},
  \href{https://doi.org/10.12942/lrr-2013-6}{\emph{Living Reviews in
  Relativity} {\bfseries 16} (Sep, 2013) 6},
  [\href{https://arxiv.org/abs/1206.1225}{{\ttfamily 1206.1225}}].

\bibitem{DES:2016}
{Dark Energy Survey Collaboration}, T.~{Abbott}, F.~B. {Abdalla},
  J.~{Aleksi{\'c}}, S.~{Allam}, A.~{Amara} et~al., \emph{{The Dark Energy
  Survey: more than dark energy - an overview}},
  \href{https://doi.org/10.1093/mnras/stw641}{\emph{\mnras} {\bfseries 460}
  (Aug., 2016) 1270--1299}, [\href{https://arxiv.org/abs/1601.00329}{{\ttfamily
  1601.00329}}].

\bibitem{HSC:2018}
H.~{Aihara}, N.~{Arimoto}, R.~{Armstrong}, S.~{Arnouts}, N.~A. {Bahcall},
  S.~{Bickerton} et~al., \emph{{The Hyper Suprime-Cam SSP Survey: Overview and
  survey design}}, \href{https://doi.org/10.1093/pasj/psx066}{\emph{\pasj}
  {\bfseries 70} (Jan., 2018) S4},
  [\href{https://arxiv.org/abs/1704.05858}{{\ttfamily 1704.05858}}].

\bibitem{LSST:2019}
Z.~{Ivezi{\'{c}}}, S.~M. {Kahn}, J.~A. {Tyson}, B.~{Abel}, E.~{Acosta},
  R.~{Allsman} et~al., \emph{{LSST: From Science Drivers to Reference Design
  and Anticipated Data Products}},
  \href{https://doi.org/10.3847/1538-4357/ab042c}{\emph{\apj} {\bfseries 873}
  (Mar., 2019) 111}, [\href{https://arxiv.org/abs/0805.2366}{{\ttfamily
  0805.2366}}].

\bibitem{SKA:2020}
A.~{Weltman}, P.~{Bull}, S.~{Camera}, K.~{Kelley}, H.~{Padmanabhan},
  J.~{Pritchard} et~al., \emph{{Fundamental physics with the Square Kilometre
  Array}}, \href{https://doi.org/10.1017/pasa.2019.42}{\emph{\pasa} {\bfseries
  37} (Jan., 2020) e002}, [\href{https://arxiv.org/abs/1810.02680}{{\ttfamily
  1810.02680}}].

\bibitem{KiDS:2020}
C.~{Heymans}, T.~{Tr{\"o}ster}, M.~{Asgari}, C.~{Blake}, H.~{Hildebrandt},
  B.~{Joachimi} et~al., \emph{{KiDS-1000 Cosmology: Multi-probe weak
  gravitational lensing and spectroscopic galaxy clustering constraints}},
  {\emph{arXiv e-prints} (July, 2020) arXiv:2007.15632},
  [\href{https://arxiv.org/abs/2007.15632}{{\ttfamily 2007.15632}}].

\bibitem{Schmidt:2015}
F.~{Schmidt}, N.~E. {Chisari} and C.~{Dvorkin}, \emph{{Imprint of inflation on
  galaxy shape correlations}},
  \href{https://doi.org/10.1088/1475-7516/2015/10/032}{\emph{\jcap} {\bfseries
  2015} (Oct., 2015) 032}, [\href{https://arxiv.org/abs/1506.02671}{{\ttfamily
  1506.02671}}].

\bibitem{chisari/etal:2016}
N.~E. {Chisari}, C.~{Dvorkin}, F.~{Schmidt} and D.~N. {Spergel},
  \emph{{Multitracing anisotropic non-Gaussianity with galaxy shapes}},
  \href{https://doi.org/10.1103/PhysRevD.94.123507}{\emph{\prd} {\bfseries 94}
  (Dec., 2016) 123507}, [\href{https://arxiv.org/abs/1607.05232}{{\ttfamily
  1607.05232}}].

\bibitem{kogai/etal}
K.~{Kogai}, K.~{Akitsu}, F.~{Schmidt} and Y.~{Urakawa}, \emph{{Galaxy imaging
  surveys as spin-sensitive detector for cosmological colliders}},
  \href{https://doi.org/10.1088/1475-7516/2021/03/060}{\emph{\jcap} {\bfseries
  2021} (Mar., 2021) 060}, [\href{https://arxiv.org/abs/2009.05517}{{\ttfamily
  2009.05517}}].

\bibitem{Chisari:2013}
N.~E. {Chisari} and C.~{Dvorkin}, \emph{{Cosmological information in the
  intrinsic alignments of luminous red galaxies}},
  \href{https://doi.org/10.1088/1475-7516/2013/12/029}{\emph{\jcap} {\bfseries
  2013} (Dec., 2013) 029}, [\href{https://arxiv.org/abs/1308.5972}{{\ttfamily
  1308.5972}}].

\bibitem{2012PhRvD..86h3513S}
F.~{Schmidt} and D.~{Jeong}, \emph{{Large-scale structure with gravitational
  waves. II. Shear}},
  \href{https://doi.org/10.1103/PhysRevD.86.083513}{\emph{\prd} {\bfseries 86}
  (Oct., 2012) 083513}, [\href{https://arxiv.org/abs/1205.1514}{{\ttfamily
  1205.1514}}].

\bibitem{chisari/etal:2014}
N.~E. {Chisari}, C.~{Dvorkin} and F.~{Schmidt}, \emph{{Can weak lensing surveys
  confirm BICEP2?}},
  \href{https://doi.org/10.1103/PhysRevD.90.043527}{\emph{\prd} {\bfseries 90}
  (Aug., 2014) 043527}, [\href{https://arxiv.org/abs/1406.4871}{{\ttfamily
  1406.4871}}].

\bibitem{Larsen:2016}
P.~{Larsen} and A.~{Challinor}, \emph{{Intrinsic alignment contamination to CMB
  lensing-galaxy weak lensing correlations from tidal torquing}},
  \href{https://doi.org/10.1093/mnras/stw1645}{\emph{\mnras} {\bfseries 461}
  (Oct., 2016) 4343--4352}, [\href{https://arxiv.org/abs/1510.02617}{{\ttfamily
  1510.02617}}].

\bibitem{2009IJMPD..18..173S}
B.~M. {Sch{\"a}fer}, \emph{{Galactic Angular Momenta and Angular Momentum
  Correlations in the Cosmological Large-Scale Structure}},
  \href{https://doi.org/10.1142/S0218271809014388}{\emph{International Journal
  of Modern Physics D} {\bfseries 18} (Jan., 2009) 173--222},
  [\href{https://arxiv.org/abs/0808.0203}{{\ttfamily 0808.0203}}].

\bibitem{1955MNRAS.115....3S}
D.~W. {Sciama}, \emph{{On the formation of galaxies in a steady state
  universe}}, \href{https://doi.org/10.1093/mnras/115.1.3}{\emph{\mnras}
  {\bfseries 115} (Jan., 1955) 3--14}.

\bibitem{1969ApJ...155..393P}
P.~J.~E. {Peebles}, \emph{{Origin of the Angular Momentum of Galaxies}},
  \href{https://doi.org/10.1086/149876}{\emph{\apj} {\bfseries 155} (Feb.,
  1969) 393}.

\bibitem{1970Ap......6..320D}
A.~G. {Doroshkevich}, \emph{{Spatial structure of perturbations and origin of
  galactic rotation in fluctuation theory}},
  \href{https://doi.org/10.1007/BF01001625}{\emph{Astrophysics} {\bfseries 6}
  (Oct., 1970) 320--330}.

\bibitem{1984ApJ...286...38W}
S.~D.~M. {White}, \emph{{Angular momentum growth in protogalaxies}},
  \href{https://doi.org/10.1086/162573}{\emph{\apj} {\bfseries 286} (Nov.,
  1984) 38--41}.

\bibitem{Catelan:2001}
P.~{Catelan} and C.~{Porciani}, \emph{{Correlations of cosmic tidal fields}},
  \href{https://doi.org/10.1046/j.1365-8711.2001.04250.x}{\emph{\mnras}
  {\bfseries 323} (May, 2001) 713--717},
  [\href{https://arxiv.org/abs/astro-ph/0012082}{{\ttfamily
  astro-ph/0012082}}].

\bibitem{Hirata:2004}
C.~M. {Hirata} and U.~{Seljak}, \emph{{Intrinsic alignment-lensing interference
  as a contaminant of cosmic shear}},
  \href{https://doi.org/10.1103/PhysRevD.70.063526}{\emph{\prd} {\bfseries 70}
  (Sept., 2004) 063526},
  [\href{https://arxiv.org/abs/astro-ph/0406275}{{\ttfamily
  astro-ph/0406275}}].

\bibitem{vlah/etal}
Z.~{Vlah}, N.~E. {Chisari} and F.~{Schmidt}, \emph{{An EFT description of
  galaxy intrinsic alignments}},
  \href{https://doi.org/10.1088/1475-7516/2020/01/025}{\emph{\jcap} {\bfseries
  2020} (Jan., 2020) 025}, [\href{https://arxiv.org/abs/1910.08085}{{\ttfamily
  1910.08085}}].

\bibitem{2015JCAP...08..015B}
J.~{Blazek}, Z.~{Vlah} and U.~{Seljak}, \emph{{Tidal alignment of galaxies}},
  \href{https://doi.org/10.1088/1475-7516/2015/08/015}{\emph{\jcap} {\bfseries
  2015} (Aug., 2015) 015}, [\href{https://arxiv.org/abs/1504.02510}{{\ttfamily
  1504.02510}}].

\bibitem{2016MNRAS.461.2702C}
N.~{Chisari}, C.~{Laigle}, S.~{Codis}, Y.~{Dubois}, J.~{Devriendt}, L.~{Miller}
  et~al., \emph{{Redshift and luminosity evolution of the intrinsic alignments
  of galaxies in Horizon-AGN}},
  \href{https://doi.org/10.1093/mnras/stw1409}{\emph{\mnras} {\bfseries 461}
  (Sept., 2016) 2702--2721},
  [\href{https://arxiv.org/abs/1602.08373}{{\ttfamily 1602.08373}}].

\bibitem{2013MNRAS.436..819J}
B.~{Joachimi}, E.~{Semboloni}, S.~{Hilbert}, P.~E. {Bett}, J.~{Hartlap},
  H.~{Hoekstra} et~al., \emph{{Intrinsic galaxy shapes and alignments - II.
  Modelling the intrinsic alignment contamination of weak lensing surveys}},
  \href{https://doi.org/10.1093/mnras/stt1618}{\emph{\mnras} {\bfseries 436}
  (Nov., 2013) 819--838}, [\href{https://arxiv.org/abs/1305.5791}{{\ttfamily
  1305.5791}}].

\bibitem{2013MNRAS.432.2433H}
C.~{Heymans}, E.~{Grocutt}, A.~{Heavens}, M.~{Kilbinger}, T.~D. {Kitching},
  F.~{Simpson} et~al., \emph{{CFHTLenS tomographic weak lensing cosmological
  parameter constraints: Mitigating the impact of intrinsic galaxy
  alignments}}, \href{https://doi.org/10.1093/mnras/stt601}{\emph{\mnras}
  {\bfseries 432} (July, 2013) 2433--2453},
  [\href{https://arxiv.org/abs/1303.1808}{{\ttfamily 1303.1808}}].

\bibitem{2019A&A...624A..30J}
H.~{Johnston}, C.~{Georgiou}, B.~{Joachimi}, H.~{Hoekstra}, N.~E. {Chisari},
  D.~{Farrow} et~al., \emph{{KiDS+GAMA: Intrinsic alignment model constraints
  for current and future weak lensing cosmology}},
  \href{https://doi.org/10.1051/0004-6361/201834714}{\emph{\aap} {\bfseries
  624} (Apr., 2019) A30}, [\href{https://arxiv.org/abs/1811.09598}{{\ttfamily
  1811.09598}}].

\bibitem{2021MNRAS.508..637S}
S.~{Samuroff}, R.~{Mandelbaum} and J.~{Blazek}, \emph{{Advances in constraining
  intrinsic alignment models with hydrodynamic simulations}},
  \href{https://doi.org/10.1093/mnras/stab2520}{\emph{\mnras} {\bfseries 508}
  (Nov., 2021) 637--664}, [\href{https://arxiv.org/abs/2009.10735}{{\ttfamily
  2009.10735}}].

\bibitem{2019MNRAS.489.5453S}
S.~{Samuroff}, J.~{Blazek}, M.~A. {Troxel}, N.~{MacCrann}, E.~{Krause}, C.~D.
  {Leonard} et~al., \emph{{Dark Energy Survey Year 1 results: constraints on
  intrinsic alignments and their colour dependence from galaxy clustering and
  weak lensing}}, \href{https://doi.org/10.1093/mnras/stz2197}{\emph{\mnras}
  {\bfseries 489} (Nov., 2019) 5453--5482},
  [\href{https://arxiv.org/abs/1811.06989}{{\ttfamily 1811.06989}}].

\bibitem{2010MNRAS.406...60J}
J.~{Jasche}, F.~S. {Kitaura}, B.~D. {Wandelt} and T.~A. {En{\ss}lin},
  \emph{{Bayesian power-spectrum inference for large-scale structure data}},
  \href{https://doi.org/10.1111/j.1365-2966.2010.16610.x}{\emph{\mnras}
  {\bfseries 406} (July, 2010) 60--85},
  [\href{https://arxiv.org/abs/0911.2493}{{\ttfamily 0911.2493}}].

\bibitem{Jasche:2013}
J.~{Jasche} and B.~D. {Wandelt}, \emph{{Bayesian physical reconstruction of
  initial conditions from large-scale structure surveys}},
  \href{https://doi.org/10.1093/mnras/stt449}{\emph{\mnras} {\bfseries 432}
  (June, 2013) 894--913}, [\href{https://arxiv.org/abs/1203.3639}{{\ttfamily
  1203.3639}}].

\bibitem{2015MNRAS.447.1204J}
J.~{Jasche} and G.~{Lavaux}, \emph{{Matrix-free large-scale Bayesian inference
  in cosmology}}, \href{https://doi.org/10.1093/mnras/stu2479}{\emph{\mnras}
  {\bfseries 447} (Feb., 2015) 1204--1212},
  [\href{https://arxiv.org/abs/1402.1763}{{\ttfamily 1402.1763}}].

\bibitem{Lavaux:2019}
G.~{Lavaux}, J.~{Jasche} and F.~{Leclercq}, \emph{{Systematic-free inference of
  the cosmic matter density field from SDSS3-BOSS data}}, {\emph{arXiv
  e-prints} (Sept., 2019) arXiv:1909.06396},
  [\href{https://arxiv.org/abs/1909.06396}{{\ttfamily 1909.06396}}].

\bibitem{Jasche:2019}
J.~{Jasche} and G.~{Lavaux}, \emph{{Physical Bayesian modelling of the
  non-linear matter distribution: New insights into the nearby universe}},
  \href{https://doi.org/10.1051/0004-6361/201833710}{\emph{\aap} {\bfseries
  625} (May, 2019) A64}, [\href{https://arxiv.org/abs/1806.11117}{{\ttfamily
  1806.11117}}].

\bibitem{Reyes:2012}
R.~{Reyes}, R.~{Mandelbaum}, J.~E. {Gunn}, R.~{Nakajima}, U.~{Seljak} and C.~M.
  {Hirata}, \emph{{Optical-to-virial velocity ratios of local disc galaxies
  from combined kinematics and galaxy-galaxy lensing}},
  \href{https://doi.org/10.1111/j.1365-2966.2012.21472.x}{\emph{\mnras}
  {\bfseries 425} (Oct., 2012) 2610--2640},
  [\href{https://arxiv.org/abs/1110.4107}{{\ttfamily 1110.4107}}].

\bibitem{2013AJ....146...32S}
S.~A. {Smee}, J.~E. {Gunn}, A.~{Uomoto}, N.~{Roe}, D.~{Schlegel}, C.~M.
  {Rockosi} et~al., \emph{{The Multi-object, Fiber-fed Spectrographs for the
  Sloan Digital Sky Survey and the Baryon Oscillation Spectroscopic Survey}},
  \href{https://doi.org/10.1088/0004-6256/146/2/32}{\emph{\aj} {\bfseries 146}
  (Aug., 2013) 32}, [\href{https://arxiv.org/abs/1208.2233}{{\ttfamily
  1208.2233}}].

\bibitem{2015ApJS..219...12A}
S.~{Alam}, F.~D. {Albareti}, C.~{Allende Prieto}, F.~{Anders}, S.~F.
  {Anderson}, T.~{Anderton} et~al., \emph{{The Eleventh and Twelfth Data
  Releases of the Sloan Digital Sky Survey: Final Data from SDSS-III}},
  \href{https://doi.org/10.1088/0067-0049/219/1/12}{\emph{\apjs} {\bfseries
  219} (July, 2015) 12}, [\href{https://arxiv.org/abs/1501.00963}{{\ttfamily
  1501.00963}}].

\bibitem{2006AJ....131.2332G}
J.~E. {Gunn}, W.~A. {Siegmund}, E.~J. {Mannery}, R.~E. {Owen}, C.~L. {Hull},
  R.~F. {Leger} et~al., \emph{{The 2.5 m Telescope of the Sloan Digital Sky
  Survey}}, \href{https://doi.org/10.1086/500975}{\emph{\aj} {\bfseries 131}
  (Apr., 2006) 2332--2359},
  [\href{https://arxiv.org/abs/astro-ph/0602326}{{\ttfamily
  astro-ph/0602326}}].

\bibitem{2013AJ....145...10D}
K.~S. {Dawson}, D.~J. {Schlegel}, C.~P. {Ahn}, S.~F. {Anderson},
  {\'E}.~{Aubourg}, S.~{Bailey} et~al., \emph{{The Baryon Oscillation
  Spectroscopic Survey of SDSS-III}},
  \href{https://doi.org/10.1088/0004-6256/145/1/10}{\emph{\aj} {\bfseries 145}
  (Jan., 2013) 10}, [\href{https://arxiv.org/abs/1208.0022}{{\ttfamily
  1208.0022}}].

\bibitem{2011AJ....142...72E}
D.~J. {Eisenstein}, D.~H. {Weinberg}, E.~{Agol}, H.~{Aihara}, C.~{Allende
  Prieto}, S.~F. {Anderson} et~al., \emph{{SDSS-III: Massive Spectroscopic
  Surveys of the Distant Universe, the Milky Way, and Extra-Solar Planetary
  Systems}}, \href{https://doi.org/10.1088/0004-6256/142/3/72}{\emph{\aj}
  {\bfseries 142} (Sept., 2011) 72},
  [\href{https://arxiv.org/abs/1101.1529}{{\ttfamily 1101.1529}}].

\bibitem{Schmidt:2012}
F.~{Schmidt} and D.~{Jeong}, \emph{{Cosmic rulers}},
  \href{https://doi.org/10.1103/PhysRevD.86.083527}{\emph{\prd} {\bfseries 86}
  (Oct., 2012) 083527}, [\href{https://arxiv.org/abs/1204.3625}{{\ttfamily
  1204.3625}}].

\bibitem{2002MNRAS.335..909B}
A.~J. {Barber}, \emph{{The redshift and scale dependence of the cosmic shear
  signal from numerical simulations}},
  \href{https://doi.org/10.1046/j.1365-8711.2002.05673.x}{\emph{\mnras}
  {\bfseries 335} (Oct., 2002) 909--917},
  [\href{https://arxiv.org/abs/astro-ph/0108273}{{\ttfamily
  astro-ph/0108273}}].

\bibitem{2016MNRAS.456..207K}
E.~{Krause}, T.~{Eifler} and J.~{Blazek}, \emph{{The impact of intrinsic
  alignment on current and future cosmic shear surveys}},
  \href{https://doi.org/10.1093/mnras/stv2615}{\emph{\mnras} {\bfseries 456}
  (Feb., 2016) 207--222}, [\href{https://arxiv.org/abs/1506.08730}{{\ttfamily
  1506.08730}}].

\bibitem{Nguyen:2020a}
N.-M. {Nguyen}, J.~{Jasche}, G.~{Lavaux} and F.~{Schmidt}, \emph{{Taking
  measurements of the kinematic Sunyaev-Zel'dovich effect forward: including
  uncertainties from velocity reconstruction with forward modeling}},
  \href{https://doi.org/10.1088/1475-7516/2020/12/011}{\emph{\jcap} {\bfseries
  2020} (12, 2020) 011--011},
  [\href{https://arxiv.org/abs/2007.13721}{{\ttfamily 2007.13721}}].

\bibitem{Blackwell:1947}
D.~Blackwell, \emph{Conditional expectation and unbiased sequential
  estimation}, \href{https://doi.org/10.1214/aoms/1177730497}{\emph{Ann. Math.
  Statist.} {\bfseries 18} (03, 1947) 105--110}.

\bibitem{Rao:1992}
C.~R. Rao, \emph{Information and the Accuracy Attainable in the Estimation of
  Statistical Parameters}, pp.~235--247.
\newblock Springer New York, New York, NY, 1992.
\newblock \href{https://doi.org/10.1007/978-1-4612-0919-5_16}{DOI}.

\bibitem{4767596}
S.~Geman and D.~Geman, \emph{Stochastic relaxation, gibbs distributions, and
  the bayesian restoration of images},
  \href{https://doi.org/10.1109/TPAMI.1984.4767596}{\emph{IEEE Transactions on
  Pattern Analysis and Machine Intelligence} {\bfseries PAMI-6} (1984)
  721--741}.

\bibitem{10.2307/2289776}
A.~E. Gelfand and A.~F.~M. Smith, \emph{Sampling-based approaches to
  calculating marginal densities}, {\emph{Journal of the American Statistical
  Association} {\bfseries 85} (1990) 398--409}.

\bibitem{2021MNRAS.504.4312G}
M.~{Gatti}, E.~{Sheldon}, A.~{Amon}, M.~{Becker}, M.~{Troxel}, A.~{Choi}
  et~al., \emph{{Dark energy survey year 3 results: weak lensing shape
  catalogue}}, \href{https://doi.org/10.1093/mnras/stab918}{\emph{\mnras}
  {\bfseries 504} (July, 2021) 4312--4336},
  [\href{https://arxiv.org/abs/2011.03408}{{\ttfamily 2011.03408}}].

\bibitem{2018MNRAS.480.3962C}
N.~E. {Chisari}, M.~L.~A. {Richardson}, J.~{Devriendt}, Y.~{Dubois},
  A.~{Schneider}, A.~M.~C. {Le Brun} et~al., \emph{{The impact of baryons on
  the matter power spectrum from the Horizon-AGN cosmological hydrodynamical
  simulation}}, \href{https://doi.org/10.1093/mnras/sty2093}{\emph{\mnras}
  {\bfseries 480} (Nov., 2018) 3962--3977},
  [\href{https://arxiv.org/abs/1801.08559}{{\ttfamily 1801.08559}}].

\bibitem{2019JCAP...03..020S}
A.~{Schneider}, R.~{Teyssier}, J.~{Stadel}, N.~E. {Chisari}, A.~M.~C. {Le
  Brun}, A.~{Amara} et~al., \emph{{Quantifying baryon effects on the matter
  power spectrum and the weak lensing shear correlation}},
  \href{https://doi.org/10.1088/1475-7516/2019/03/020}{\emph{\jcap} {\bfseries
  2019} (Mar., 2019) 020}, [\href{https://arxiv.org/abs/1810.08629}{{\ttfamily
  1810.08629}}].

\bibitem{2021MNRAS.503.3796D}
C.~{Doux}, C.~{Chang}, B.~{Jain}, J.~{Blazek}, H.~{Camacho}, X.~{Fang} et~al.,
  \emph{{Consistency of cosmic shear analyses in harmonic and real space}},
  \href{https://doi.org/10.1093/mnras/stab661}{\emph{\mnras} {\bfseries 503}
  (May, 2021) 3796--3817}, [\href{https://arxiv.org/abs/2011.06469}{{\ttfamily
  2011.06469}}].

\bibitem{2021MNRAS.501.2983F}
M.~C. {Fortuna}, H.~{Hoekstra}, B.~{Joachimi}, H.~{Johnston}, N.~E. {Chisari},
  C.~{Georgiou} et~al., \emph{{The halo model as a versatile tool to predict
  intrinsic alignments}},
  \href{https://doi.org/10.1093/mnras/staa3802}{\emph{\mnras} {\bfseries 501}
  (Feb., 2021) 2983--3002}, [\href{https://arxiv.org/abs/2003.02700}{{\ttfamily
  2003.02700}}].

\bibitem{2008ApJ...681..798L}
J.~{Lee} and U.-L. {Pen}, \emph{{The Nonlinear Evolution of Galaxy Intrinsic
  Alignments}}, \href{https://doi.org/10.1086/588646}{\emph{\apj} {\bfseries
  681} (July, 2008) 798--805},
  [\href{https://arxiv.org/abs/0707.1690}{{\ttfamily 0707.1690}}].

\bibitem{2021MNRAS.502.3035P}
N.~{Porqueres}, A.~{Heavens}, D.~{Mortlock} and G.~{Lavaux}, \emph{{Bayesian
  forward modelling of cosmic shear data}},
  \href{https://doi.org/10.1093/mnras/stab204}{\emph{\mnras} {\bfseries 502}
  (Apr., 2021) 3035--3044}, [\href{https://arxiv.org/abs/2011.07722}{{\ttfamily
  2011.07722}}].

\bibitem{2021arXiv210514699F}
P.~{Fiedorowicz}, E.~{Rozo}, S.~S. {Boruah}, C.~{Chang} and M.~{Gatti},
  \emph{{$\texttt{KaRMMa}$ -- Kappa Reconstruction for Mass Mapping}},
  {\emph{arXiv e-prints} (May, 2021) arXiv:2105.14699},
  [\href{https://arxiv.org/abs/2105.14699}{{\ttfamily 2105.14699}}].

\bibitem{2021arXiv210914297T}
M.~{Tonegawa} and T.~{Okumura}, \emph{{First Evidence of Intrinsic Alignments
  of Red Galaxies at $z > 1$: Cross-correlation between CFHTLenS and FastSound
  Samples}}, {\emph{arXiv e-prints} (Sept., 2021) arXiv:2109.14297},
  [\href{https://arxiv.org/abs/2109.14297}{{\ttfamily 2109.14297}}].

\bibitem{2019A&A...624A.115P}
N.~{Porqueres}, D.~{Kodi Ramanah}, J.~{Jasche} and G.~{Lavaux}, \emph{{Explicit
  Bayesian treatment of unknown foreground contaminations in galaxy surveys}},
  \href{https://doi.org/10.1051/0004-6361/201834844}{\emph{\aap} {\bfseries
  624} (Apr., 2019) A115}, [\href{https://arxiv.org/abs/1812.05113}{{\ttfamily
  1812.05113}}].

\bibitem{2017A&A...606A..37J}
J.~{Jasche} and G.~{Lavaux}, \emph{{Bayesian power spectrum inference with
  foreground and target contamination treatment}},
  \href{https://doi.org/10.1051/0004-6361/201730909}{\emph{\aap} {\bfseries
  606} (Oct., 2017) A37}, [\href{https://arxiv.org/abs/1706.08971}{{\ttfamily
  1706.08971}}].

\bibitem{2019A&A...621A..69R}
D.~K. {Ramanah}, G.~{Lavaux}, J.~{Jasche} and B.~D. {Wandelt},
  \emph{{Cosmological inference from Bayesian forward modelling of deep galaxy
  redshift surveys}},
  \href{https://doi.org/10.1051/0004-6361/201834117}{\emph{\aap} {\bfseries
  621} (Jan., 2019) A69}, [\href{https://arxiv.org/abs/1808.07496}{{\ttfamily
  1808.07496}}].

\bibitem{2022arXiv220308838A}
A.~{Andrews}, J.~{Jasche}, G.~{Lavaux} and F.~{Schmidt}, \emph{{Bayesian
  field-level inference of primordial non-Gaussianity using next-generation
  galaxy surveys}}, {\emph{arXiv e-prints} (Mar., 2022) arXiv:2203.08838},
  [\href{https://arxiv.org/abs/2203.08838}{{\ttfamily 2203.08838}}].

\bibitem{2016PhRvD..94l3507C}
N.~E. {Chisari}, C.~{Dvorkin}, F.~{Schmidt} and D.~N. {Spergel},
  \emph{{Multitracing anisotropic non-Gaussianity with galaxy shapes}},
  \href{https://doi.org/10.1103/PhysRevD.94.123507}{\emph{\prd} {\bfseries 94}
  (Dec., 2016) 123507}, [\href{https://arxiv.org/abs/1607.05232}{{\ttfamily
  1607.05232}}].

\bibitem{1992StaSc...7..457G}
A.~{Gelman} and D.~B. {Rubin}, \emph{{Inference from Iterative Simulation Using
  Multiple Sequences}},
  \href{https://doi.org/10.1214/ss/1177011136}{\emph{Statistical Science}
  {\bfseries 7} (Jan., 1992) 457--472}.

\bibitem{10.2307/2280095}
F.~J. Massey, \emph{The kolmogorov-smirnov test for goodness of fit},
  {\emph{Journal of the American Statistical Association} {\bfseries 46} (1951)
  68--78}.

\bibitem{10.1214/aoms/1177729694}
S.~Kullback and R.~A. Leibler, \emph{{On Information and Sufficiency}},
  \href{https://doi.org/10.1214/aoms/1177729694}{\emph{The Annals of
  Mathematical Statistics} {\bfseries 22} (1951) 79 -- 86}.

\bibitem{10.2307/2331554}
Student, \emph{The probable error of a mean}, {\emph{Biometrika} {\bfseries 6}
  (1908) 1--25}.

\end{thebibliography}\endgroup

\label{lastpage}
\end{document}